\documentclass[twocolumn,journal]{IEEEtran}
\usepackage[T1]{fontenc}
\usepackage[latin9]{inputenc}
\usepackage{color}
\usepackage{amsmath}
\usepackage{amssymb}
\usepackage{graphicx}
\usepackage {hyperref}
\hypersetup{
 pdfborderstyle={},pdfinfo={ Creator={},Producer={}}}

\makeatletter
\usepackage{color}
\usepackage{float}
\usepackage{breakurl}
\usepackage{subfig}

\usepackage{comment}

\AtBeginDocument{%
\let\ref\autoref
\renewcommand\equationautorefname{\@gobble}
}
\usepackage[noadjust]{cite}

\newcommand{\tlua}[1]{\textcolor{black}{#1}}
\definecolor{orange}{RGB}{255,127,0}

\author{
Thong~Nguyen\thanks{Thong~Nguyen~and~Jos\'e~Schutt-Ain\'e~are~with~the~University~of~Illinois~at~Urbana~-~Champaign,~Illinois,~USA, e-mail:~\{tnnguye3, jesa\}@illinois.edu},~\IEEEmembership{Student~Member,~IEEE},
Tianjian~Lu\thanks{Tianjian~Lu~and~Ken~Wu~are~with~Google~Inc.,~1600~Amphitheatre~Pkwy,~Mountain~View,~CA~94043,~ USA,~e-mail:~\{tianjianlu, kenzwu\}@google.com.},~\IEEEmembership{Member,~IEEE},
Ken~Wu,~\IEEEmembership{Member,~IEEE},
\and~Jos\'e~Schutt-Ain\'e,~\IEEEmembership{Member,~IEEE}
}

\usepackage{cite}

\makeatother

\begin{document}

\title{\tlua{Fast Transient Simulation of High-Speed Channels Using Recurrent
Neural Network}}
\maketitle

\begin{abstract}
Generating eye diagrams by using a circuit simulator can be very computationally intensive, especially in the presence of nonlinearities.
It often involves multiple Newton-like iterations at every time step when a SPICE-like circuit simulator handles a nonlinear system in the transient regime.
In this paper, we leverage machine learning methods, to be specific, the recurrent neural network (RNN), to generate black-box macro-models and achieve significant reduction of computation time.
Through the proposed approach, an RNN model is first trained and
then validated on a relatively short sequence generated from a circuit
simulator. Once the training completes, the RNN can be used to make
predictions on the remaining sequence in order to generate an eye
diagram. The training cost can also be amortized when the trained
RNN starts making predictions.  
Besides, the proposed approach requires
no complex circuit simulations nor substantial domain knowledge. 
We use two high-speed link examples to demonstrate that the proposed approach
provides adequate accuracy while the computation
time can be dramatically reduced.
In the high-speed link example with a PAM4 driver, the eye diagram generated by RNN models shows good agreement with that obtained from a commercial circuit simulator.
This paper also investigates the impacts of various RNN topologies, training schemes, and tunable parameters on both the accuracy and the generalization capability of an RNN model. 
It is found out that the long short-term memory (LSTM) network outperforms the vanilla RNN in terms of the accuracy in predicting transient waveforms.
\end{abstract}


\section{Introduction}

\begin{figure}[t]
\centering{}\includegraphics[scale=0.35]{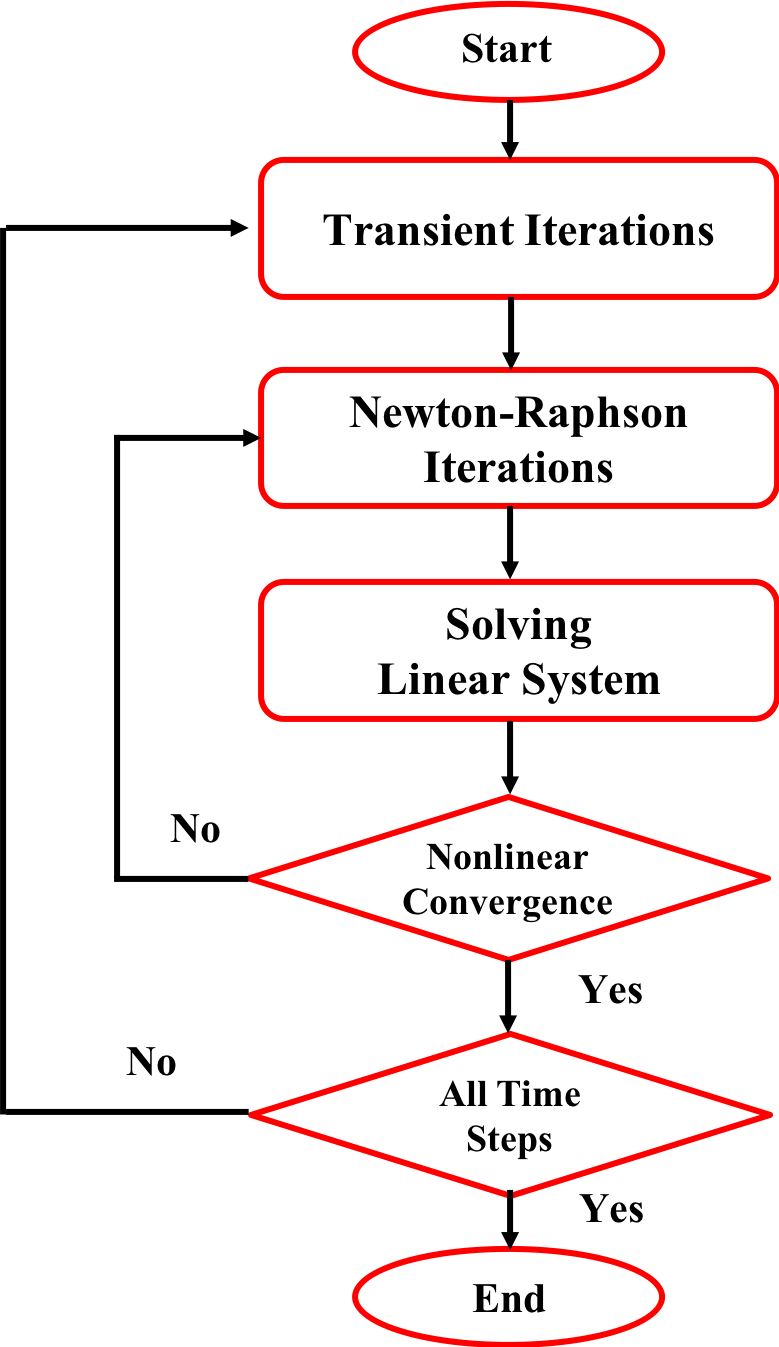}
\caption{Flow chart of a circuit simulator \cite{SPICE_DeHon}.\label{fig:flow_chart_spice}}
\end{figure}

In the area of signal integrity, eye diagrams have become important
metrics to assess the performance of a high-speed channel. In order
to generate an eye diagram, transient waveforms are first obtained
from a circuit simulator and then overlaid. Generating eye diagrams
by using a circuit simulator can be very computationally intensive,
especially in the presence of nonlinearities. As shown in \ref{fig:flow_chart_spice},
there are {often multiple} Newton-like iterations involved at every
time step when a SPICE-like circuit simulator handles a nonlinear
system in the transient regime \cite{SPICE_DeHon}. Given the size
of a practical and large-scale circuit, the runtime of a circuit simulator
on modern processors can be hours, days, or even weeks. There are
many efforts in seeking novel numerical techniques to improve the
computation efficiency of a circuit simulator. For example, people
are {using hardware accelerators including} FPGAs \cite{SPICE_DeHon}
and GPUs \cite{GPU_SPICE_Yang} to achieve the acceleration of matrix
factorization in a circuit simulator. There are also works in efficiently
generating eye diagrams, for example, using shorter bit patterns instead
of the pseudo-random bit sequence as input sources to simulate the
worst-case eye diagram \cite{EyeGen_RBWu}. In this work, we propose
taking a different route and using machine learning methods, to be
specific, recurrent neural network (RNN), to improve the efficiency
of a circuit simulator. 

Recently, many remarkable results are reported on modern time-series
techniques by using RNN in the fields such as language modeling, machine
translation, chatbot, and forecasting \cite{RNN_app_uber,RNN_app_seq2seq2,RNN_app_seq2seq,RNN_app_imgreg,RNN_app_seqgen}.
There are also a number of prior attempts in incorporating RNN into
modeling and {simulating electronic devices and systems}. For example,
researchers propose combining a NARX (nonlinear auto-regressive network)
topology with a feedforward neural network in modeling nonlinear RF
devices \cite{RNN_narx_rfckt_review}. A variant of RNN, known as
Elman RNN (ERNN), is applied in simulating digital designs \cite{RNN_ERNN1,RNN_ERNN2}.
More recently, researchers present an ERNN-based model in simulating
electrostatic discharge (ESD) \cite{RNN_ERNNzaichen}. The aforementioned
two topologies, to be specific, NARX-RNN and ERNN, will be discussed
in details in the following sections. It is also worth mentioning
that machine learning methods in general have been seen into many
applications related to electronic designs such as modeling high-speed
channels \cite{Eye_DNN_Lu,Eye_S_FNN_Nikita,Eye_Spar_FNN_Goh}, replacing
computationally expensive full-wave electromagnetic simulations \cite{NN_geom_2_S_ustrip,NN_geom_2_S_via},
and building macro-model from S-parameters \cite{SFNN,NN_poleres}.

Through the proposed approach, a modern ERNN is first trained and
then validated on a relatively short sequence generated from a circuit
simulator. Once the training completes, the RNN can be used to make
predictions on the remaining sequence in order to generate an eye
diagram. The training cost can also be amortized when the trained
RNN starts making predictions. As the time-domain waveforms are generated
from RNN through inference instead of iterations of solving linear
systems involved in a circuit simulator, it significantly improves
the computation efficiency. Besides, the proposed approach requires
no complex circuit simulations nor substantial domain knowledge. We
demonstrate {through two examples} that the proposed approach can
meet the accuracy of transistor-level simulation while {the} run
time {can be} dramatically reduced.

In this work, we also investigate the performance of ERNNs built with
different recurrent units, to be specific, vanilla recurrent {neural
network} (VRNN), long short-term memory (LSTM) {unit}, and gated
recurrent {unit} (GRU) in generating accurate eye diagrams. It is
shown that {the LSTM network outperforms VRNN in terms of both convergence
and accuracy. The numerical issue of gradient vanishing or explosion
during back propagation in the VRNN is also well resolved in the LSTM
network.} The activation function in this work is {chosen as the
rectified linear unit (ReLU)} \cite{DL_ReLU} {as it enables better
numerical stability and higher efficiency in training.} \emph{Adam}
\cite{DL_Adam} optimizer is found to stand out among investigated
optimizers such as \emph{Stochastic Gradient Descent (SGD) \cite{DL_SGD},
Root Mean Square Propagation (RMSProp)} \cite{tieleman2014rmsprop}
for fast convergence in training. \tlua{It is also shown that with
the training scheme proposed in this paper, training sample length
plays an important role to the convergence of the RNN.}

\section{Recurrent neural network\label{sec:Recurrent-neural-network}}

{Understanding RNN cannot be separated from} the feed-forward neural network (FNN), 
which consists of multiple layers of neurons. Unlike a FNN, in which the signal flows unidirectionally
from the input to the output, a RNN has, in addition, a feedback loop
from the output to the input. The FNN is a universal approximator,
which can be written as the following 
\begin{equation}
y=f_{L}\circ f_{L-1}\circ...\circ f_{1}\circ x,\label{eq:ffn}
\end{equation}
where $x\in\mathbb{R}^{n}$ and $y\in\mathbb{R}^{m}$ represent the
input and the output, respectively, $f_{l}\thinspace$ ($l=1,2...,L$)
is the weighted activation, and $\circ$ denotes the composition operation.
As a comparison, the RNN can be understood as a universal Turing machine
in the form of 
\begin{equation}
\left\{ \begin{array}{ccl}
h_{t} & \!\!\!=\!\!\! & g_{h}\left(x_{t},h_{t-1}\right)\\
y_{t} & \!\!\!=\!\!\! & g_{o}\left(h_{t}\right),
\end{array}\right.\label{eq:ernn-dynamic}
\end{equation}
where $h_{t}$ and $x_{t}$ are the hidden state and the input at
time $t$, respectively, and $g_{h}$ and $g_{o}$ are weighted activations.

Similar to that in a dynamical system, the concept of state is employed
to describe the temporal evolution of a system, the power of a RNN
in dealing with time-series tasks arises from the special variable,
namely, the hidden (internal) state $h_{t}$. In system identification,
the mappings including both $g_{h}$ and $g_{o}$ in Equation (\ref{eq:ernn-dynamic})
are learnt via a least-square alike approximation process during which
a set of pre-defined parameters are tuned. Similar models to the one
described by Equation (\ref{eq:ernn-dynamic}) can be found in \emph{autoregressive
(AR)} family, which are also very popular for time-series tasks. The
models of a AR family can often be implemented with 
\begin{equation}
y_{t}=g\left(x_{t-i},y_{t-j}\right)\ \ \ 0\leq i\leq K_{x},0\leq j<K_{y},\label{eq:ar}
\end{equation}
where $K_{x}$ and $K_{y}$ are known as the memory length of the
input and output, respectively. It can be seen from Equation (\ref{eq:ar})
that there is no explicit hidden state; instead, the feedback comes
from the delayed versions of the output. In order to differentiate
the mechanism described in Equation (\ref{eq:ar}), the RNN with explicitly
defined hidden states are often called the Elman RNN (ERNN) \cite{RNN_Elman90findingstructure}.
In this work, we use RNN to denote ERNN for simplicity. {The
term NARX-RNN and output-feedback RNN will be used interchangbly to
refer to AR-based RNN.}

\begin{figure}[ht]
\begin{centering}
\includegraphics[scale=0.7]{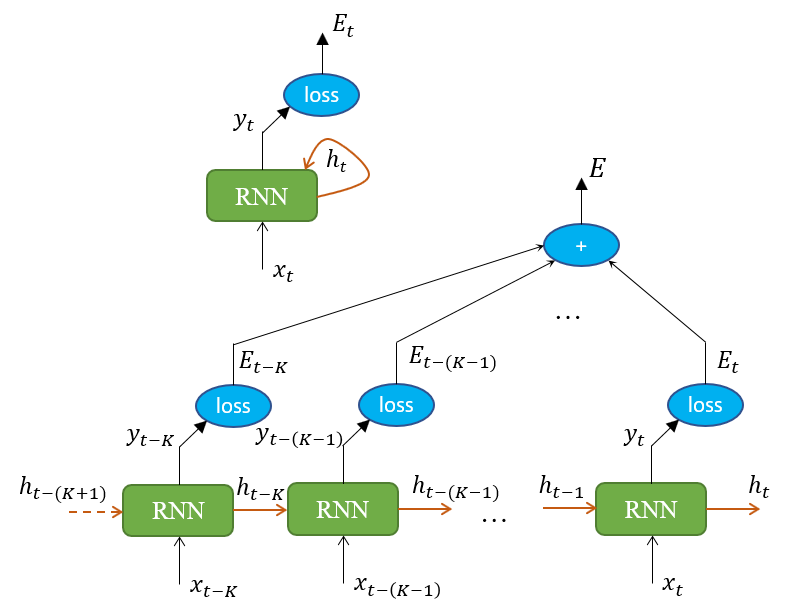} 
\par\end{centering}
\caption{An unrolled RNN with input sequence of K steps \label{fig:Unrolling-an-RNN}
with $\tilde{y}_{\tau}$ and $E_{\tau}$ representing the prediction
and the corresponding loss (error) at time step $\tau$.}
\end{figure}

It is often beneficial to unroll a RNN, which will ease the understanding
for why the learning process of a RNN could be computationally intractable
and how it is made tractable. As shown in \ref{fig:Unrolling-an-RNN},
the RNN is unrolled such that it can be fed with an input sequence
of $K$ time steps. The signal propagating through a unit in the unrolled
RNN can thus be written as 
\begin{equation}
h_{t}=\phi_{h}\left(W_{ih}x_{t}+W_{hh}h_{t-1}\right)\label{eq:ernn-states}
\end{equation}
and the output of the RNN unit is given by 
\begin{equation}
y_{t}=h_{t},\label{eq:ernn_out}
\end{equation}
where $\phi_{h}$ is the nonlinear activation function and $W$ contains
the tunable weights. It is worth mentioning that one can always add
a fully connected layer to $y_{t}$ in Equation \eqref{eq:ernn_out}
to transform it into the desired form, which is also the reason why
modern formulation of RNN takes the current state as the output.

The unrolled RNN looks likes a deep FNN (DNN), but the weights are
shared across the units over time. It is an advantage of RNN over
FNN as by unrolling the RNN, one obtains a DNN of the same number
of layers but with much fewer parameters. Unfortunately, this also
leads to disadvantages of RNN, which can be understood in the following.
The gradient of the loss $E$ at output with respect to a parameter
$\theta$ can be written as 
\begin{equation}
\frac{\partial E}{\partial\theta}=\sum_{\tau=1}^{K}\frac{\partial E_{\tau}}{\partial\theta},
\end{equation}
where 
\begin{equation}
\frac{\partial E_{\tau}}{\partial\theta}=\sum_{j=1}^{\tau}\frac{\partial E_{\tau}}{\partial\tilde{y}_{\tau}}\frac{\partial\tilde{y}_{\tau}}{\partial h_{\tau}}\frac{\partial h_{\tau}}{\partial h_{j}}\frac{\partial h_{j}}{\partial\theta}.\label{localE}
\end{equation}
The parameters in the RNN are updated through the backpropagation
of the calculated gradients. The backpropagation of the gradients
from time $\tau$ are done through all possible routes toward the
past, which is also known as backpropagation through time (BPTT).

\begin{figure}[h]
\begin{centering}
\subfloat[Sigmoid]{\begin{centering}
\includegraphics[scale=0.5]{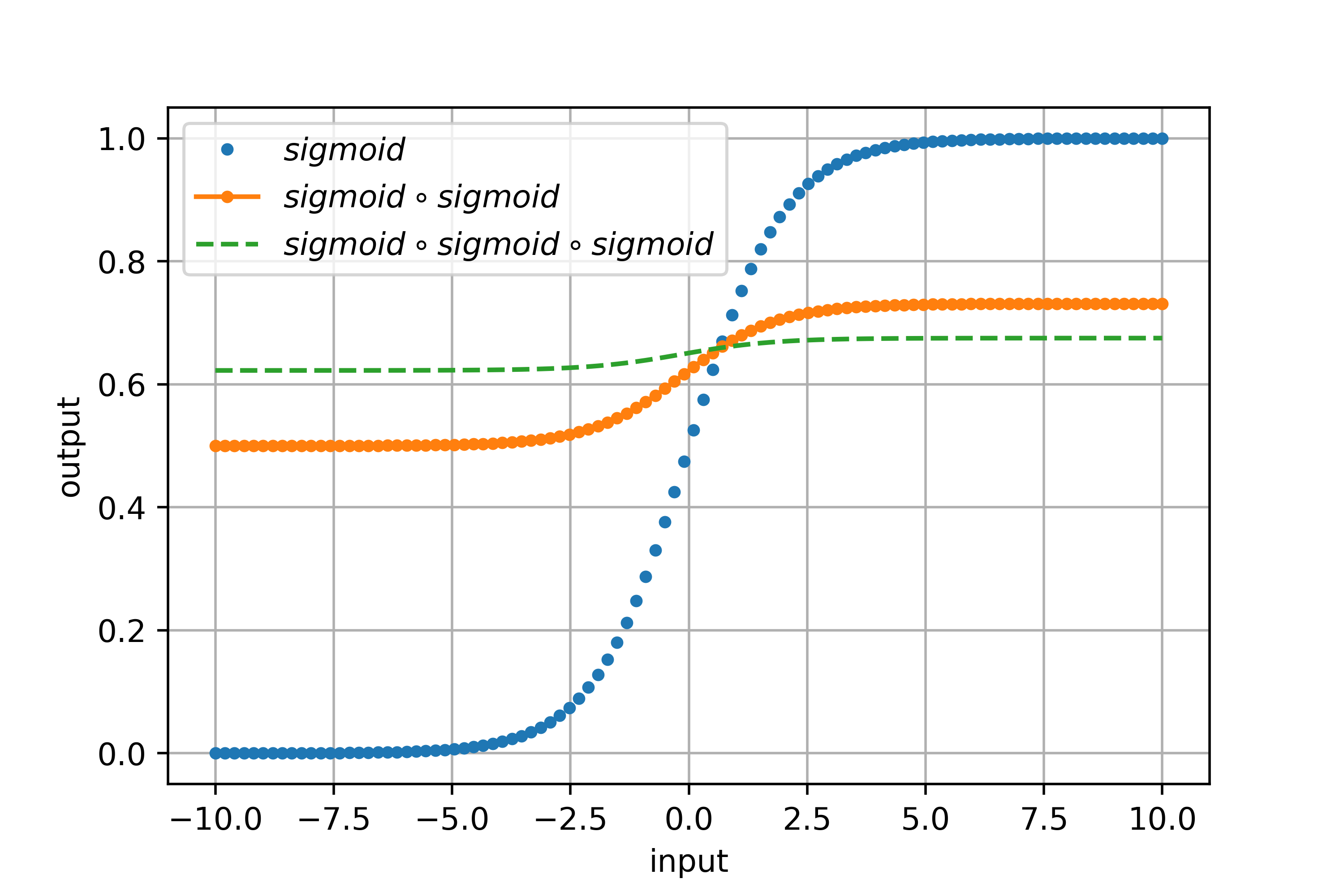} 
\par\end{centering}
}
\vspace{-0.15in}
\par\end{centering}
\begin{centering}
\subfloat[Tanh]{\begin{centering}
\includegraphics[scale=0.5]{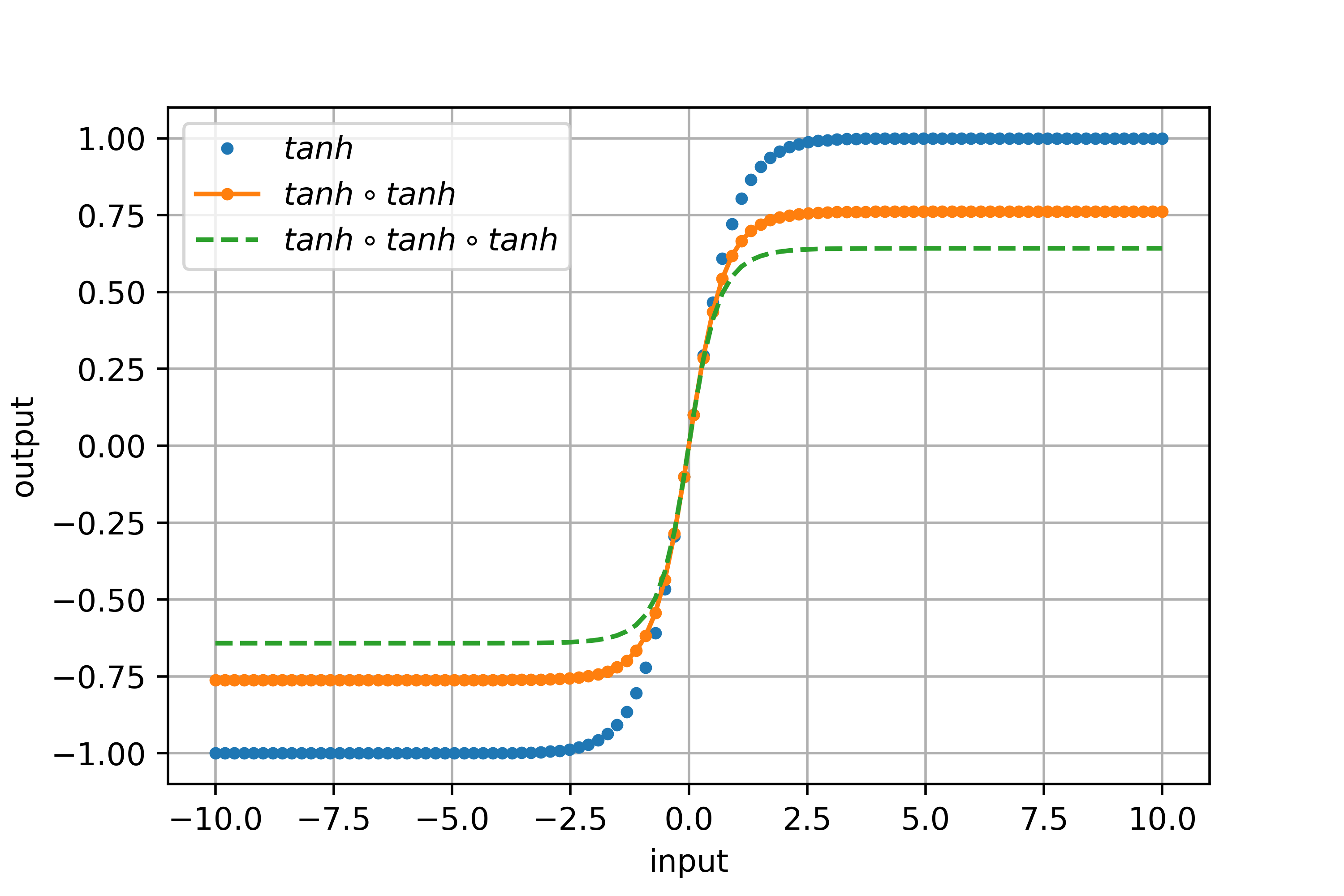} 
\par\end{centering}
}
\vspace{-0.15in}
\par\end{centering}
\begin{centering}
\subfloat[ReLU]{\begin{centering}
\includegraphics[scale=0.5]{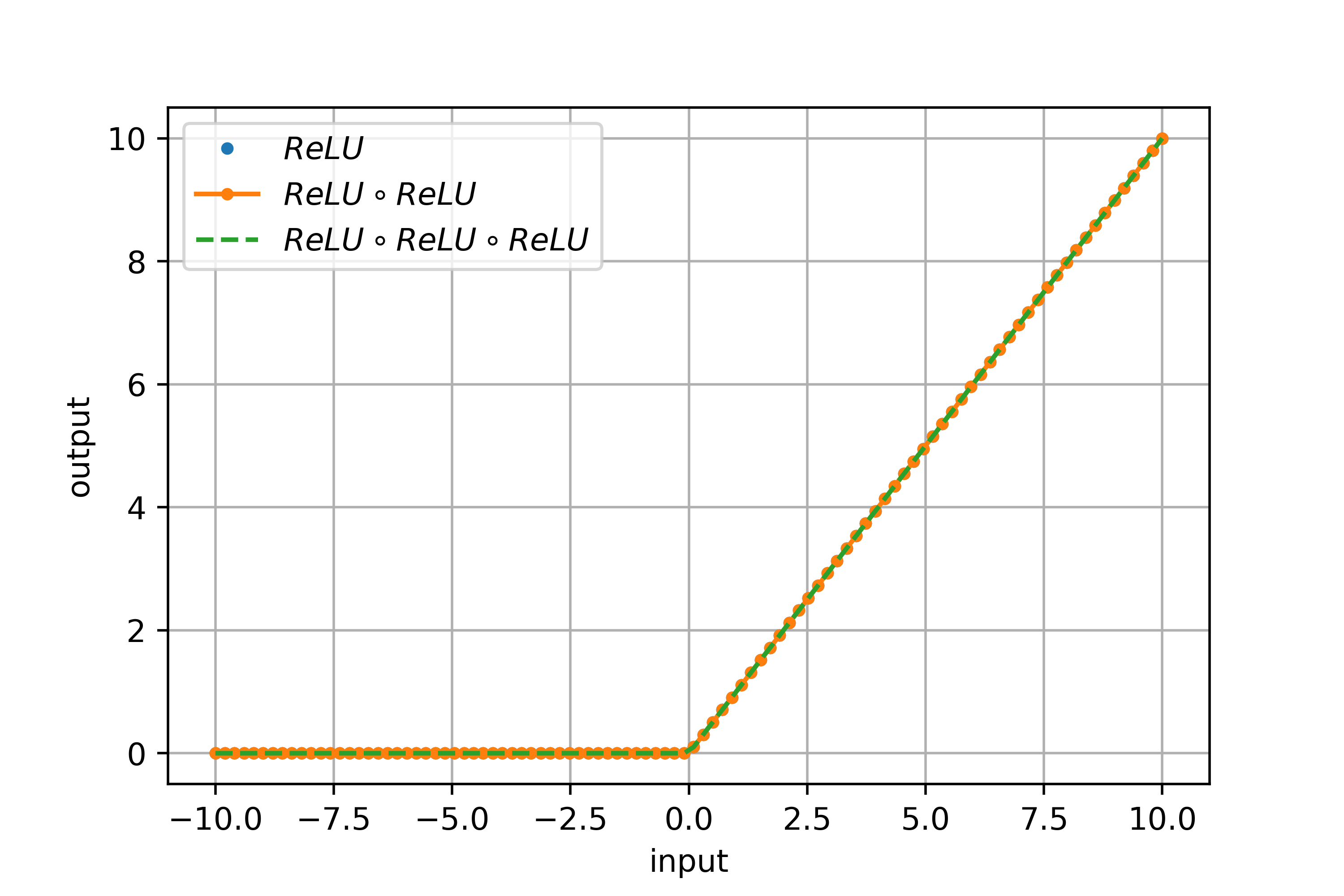} 
\par\end{centering}
}
\par\end{centering}
\caption{Multiple passes through the same activation function.\label{fig:Multiple-passes-through}}
\end{figure}

One disadvantage on BPTT is {the} computation efficiency because at any
time step $\tau$, the calculation of the loss $E_{\tau}$ depend
on all previous quantities. It can be seen that with BPTT, the longer
the sequence with which the RNN is trained, the more challenging the
computation becomes considering both the degraded convergence rate
and the increased demand on computing resources. Another numerical
issue associated with the gradients with BPTT is that as the span
of the temporal dependencies increases, the gradients tend to vanish
or explode. The Jacobian term in the gradient of the loss function,
${\displaystyle \frac{\partial h_{\tau}}{\partial h_{j}}}$ in Equation
(\ref{localE}) can be proved to be upper bounded by a geometric series
\cite{RNN_bengio_grad_in_RNN}: 
\begin{equation}
\left\Vert \frac{\partial h_{\tau}}{\partial h_{j}}\right\Vert <\gamma^{\tau-j},\label{eq:vanish-grad}
\end{equation}
where $\gamma$ is a constant {determined} by the norm of the nonlinearity
in RNN. When {the} hyperbolic tangent function $\tanh$ is chosen as the
activation function, {we have} $\gamma\!=\!1$, and for the sigmoid function,  $\gamma\!=\!0.25$
\cite{RNN_bengio_grad_in_RNN}. Therefore, the gradient either explodes
or vanishes. We can use a numerical experiment to demonstrate the
gradient vanishing and explosion. As shown in \ref{fig:Multiple-passes-through},
an input signal whose magnitude ranges from $-10$ to $10$ is passed
through various types of activation functions in multiple times. After
the third time, the signal is flattened when the sigmoid function is
taken as the activation function. Due to the vanishing of the gradients,
the sigmoid function cannot be used as the activation function in a
RNN unit. In contrast, as shown in \ref{fig:Multiple-passes-through}(c)
when $\text{ReLU}$ is taken as the activation function, the signal
remains {as} its original shape after being passed through the unit for
iterations, which is also the reason why ReLU is very popular in modern
RNN structures.

One remedy to the problem of gradient vanishing or exploding is known
as truncated backpropagation through time (TBPTT) \cite{RNN_TBPTT1,RNN_TBPTT_sutskever2013},
which is a modified version of BPTT. A TBPTT processes the sequence
one step at a time, and after every $k_{1}$ time steps, it calls
BPTT for $k_{2}$ time steps. A common configuration of TBPTT is {that} the
forward and backward processes share the same number of steps such that
$k_{1}\!=\!k_{2}$. Another remedy utilizes a more sophisticated activation
function with gating units to deal with problem of gradient vanishing
or explosion, for example, the long short-term memory (LSTM) unit
\cite{RNN_COlah_LSTM} and the gated recurrent unit (GRU) \cite{RNN_GRU}.
{Both LSTM unit and GRU} own \emph{gates}, which allow the RNN cell to \emph{forget}. 
{The working mechanism of a LSTM network is based on}

\begin{equation}
\left\{ \begin{array}{ccl}
i_{t}\!\!\!\! & = & \!\!\!\!\sigma\left(W_{ii}x_{t}+W_{hi}h_{t-1}\right)\\
f_{t}\!\!\!\! & = & \!\!\!\!\sigma\left(W_{if}x_{t}+W_{hf}h_{t-1}\right)\\
g_{t}\!\!\!\! & = & \!\!\!\!\tanh\left(W_{ig}x_{t}+W_{hg}h_{t-1}\right)\\
o_{t}\!\!\!\! & = & \!\!\!\!\sigma\left(W_{io}x_{t}+W_{ho}h_{t-1}\right)\\
c_{t}\!\!\!\! & = & \!\!\!\!f_{t}c_{t-1}+i_{t}g_{t}\\
h_{t}\!\!\!\! & = & \!\!\!\!o_{t}\tanh\left(c_{t}\right),
\end{array}\right.\label{eq:lstm}
\end{equation}
where $h_{t}$ is hidden state at time $t$, $c_{t}$ is the cell
state, {and} $i_{t},f_{t},g_{t},o_{t}$ are the input, forget, cell and
output gates respectively.

{As a comparison to a LSTM unit, there is no cell component in GRU and the working mechanism of a GRU network is as follows}
\begin{equation}
\left\{ \begin{array}{ccl}
r_{t}\!\!\!\! & = & \!\!\!\!\sigma\left(W_{ir}x_{t}+W_{hr}h_{t-1}\right)\\
z_{t}\!\!\!\! & = & \!\!\!\!\sigma\left(W_{iz}x_{t}+W_{hz}h_{t-1}\right)\\
n_{t}\!\!\!\! & = & \!\!\!\!\tanh\left(W_{in}x_{t}+r_{t}W_{hn}h_{t-1}\right)\\
h_{t}\!\!\!\! & = & \!\!\!\!\left(1-z_{t}\right)n_{t}+z_{t}h_{t-1},
\end{array}\right.\label{eq:gru}
\end{equation}
where $z_{t}$ and $r_{t}$ are {called the} \emph{update} and {the} \emph{reset} gates.
{Both $z_{t}$ and $r_{t}$ function as} control signals {within the unit}. The GRU {employs a new way
to calculate the memory $n_{t}$ by using}
the current input and the past hidden state. It can be seen that a
LSTM unit requires a more complex implementation on the gating functions
than a GRU. However, both LSTM unit and GRU are able to store
and retrieve relevant information from the past by using gating control
signal, which resolves the issue of gradient vanishing or explosion
\cite{RNN_COlah_LSTM}.

\section{Training an RNN\label{sec:Training-an-RNN}}

In this section, we will review three different {training schemes,
namely, \emph{readout}, \emph{teacher force}, and \emph{professor
force}.} The most trivial way to train an RNN is the {\emph{readout}
training as shown in \ref{fig:Readout-training}. By taking advantage
of the recurrent nature of an RNN, the \emph{readout} training takes
the output at previous time steps as the input.} The ground-truth
$\tilde{y}_{k}$ is only used in calculating loss with the corresponding
prediction $y_{k}$. The RNN is fed with what it generated, which
is also the reason it is called \emph{readout}. \emph{Readout} technique
is mostly adopted in inference, i.e. when {predictions are being
made} on the unseen data. However, training in {\emph{readout} mode
often takes longer time on convergence because the model has to make
a lot of mistakes, being penalized for many times before it eventually
learns to generate accurate predictions. Therefore, \emph{teacher
force} training is often preferred over \emph{readout}. In \emph{teacher
force} training as illustrated in \ref{fig:teacher-training}, the
ground-truth values are fed into an RNN as input. Teacher forcing
can ensure an RNN learn faster but not necessarily better.

\begin{figure}[h]
\begin{centering}
\includegraphics[scale=0.6]{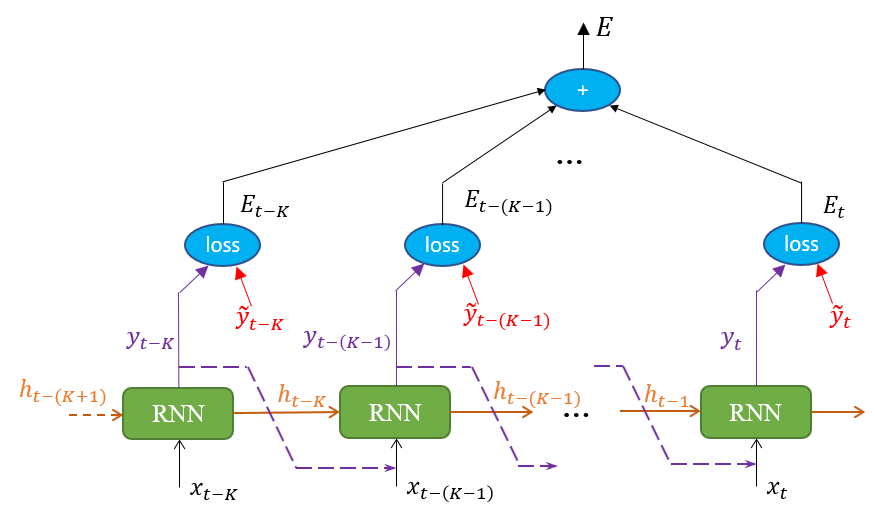} 
\par\end{centering}
\caption{Readout training.\label{fig:Readout-training}}
\end{figure}

\begin{figure}[h]
\begin{centering}
\includegraphics[scale=0.6]{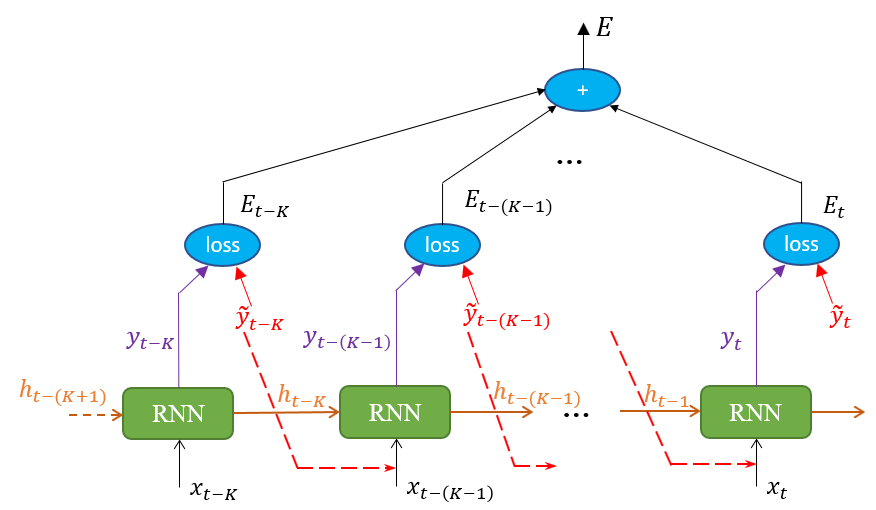} 
\par\end{centering}
\caption{Teacher force training.\label{fig:teacher-training}}
\end{figure}

{Similar to the mechanism behind overfitting, the underlying distribution
of the input data in \emph{teacher force} training may be very different
from that during its \emph{readout} mode inference. In that case,
\emph{teacher force} training may have worse performance on unseen
data comparing to its performance on the training set. To filter out
the potential bias in training, a scheduling process can be adopted
\cite{RNN_schedule-sampl}. A good analogy of the scheduling process
is the event of flipping a coin: we can imagine that a coin is flipped
every time before the previous output is fed into an RNN as the input.
The coin used in the scheduling process is biased: for the first few
training epochs, the coin is biased towards the training data distribution
such that the training is more into a \emph{teacher force} mode; as
the training evolves, the coin becomes biased towards the distribution
of the predicted data, in other words, in a \emph{readout} mode. The
scheduling technique has lead to a significant improvement on the
generalization capability of an RNN model in speech recognition \cite{RNN_coco_img_caption}.}
{In this paper, we use the \emph{teach force} training with scheduling
for the transient channel simulation example shown in \ref{sec:example}.}

{There is an another training scheme called \emph{professor force}
\cite{RNN_prof_force} technique which enforces the similar behaviors
of the network during training and test. The \emph{professor force}
training uses the concept from generative modeling, to be specific,
generative adversarial network (GAN) to fill the gap between the training
distribution and the predictive distribution, providing better generalization
performance. } }

\section{Numerical examples\label{sec:example}}

\tlua{The robustness of the proposed method using RNN to model high-speed channels will be illustrated through two different types of RNN presented in \ref{sec:Recurrent-neural-network} using a PAM2 and a PAM4 driver circuit. Different training conditions such as optimization method, memory length and recurrent cell topology etc. will be investigated.
}

\subsection{PAM2 channel simulation with output-feedback RNN (NARX-RNN)}

\begin{figure}[h]
\begin{centering}
\includegraphics[scale=0.5]{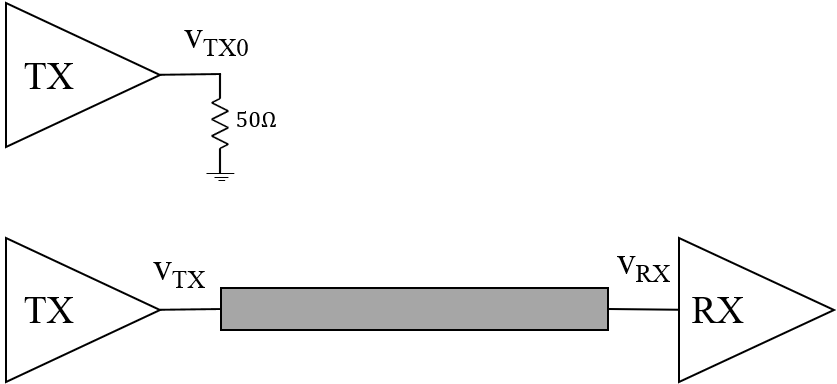} 
\par\end{centering}
\caption{{Simulation} setup for data collection.\label{fig:Measurement-setup-for}.}
\end{figure}

In this section, we illustrate the training procedures of the RNN,
with which the predictions can be made on the voltage waves arriving
at the receiver of a high-speed channel using NARX-RNN. 
\tlua{As demonstrated in Equation (\ref{eq:ar}), the NARX-RNN does not have a hidden
state explicitly defined in the model. The current output response is determined only using the current and past values of the input and the past values of the output.} The set up is shown in \ref{fig:Measurement-setup-for}:
$V_{TX0}$ is the output voltage of \tlua{the transmitter (TX)} when {it is terminated with a 50 Ohm resistor}; and $V_{TX}$ and
$V_{RX}$ are voltages at the {immediate} output of TX and the input of RX {in the presence of the channel}. 
{In this example, w}e use $V_{TX}$ and $V_{TX0}$ of the current
time step and $V_{RX}$ of the past to predict $V_{RX}$ of
the current time step.

\begin{figure}[h]
\begin{centering}
\includegraphics[scale=0.5]{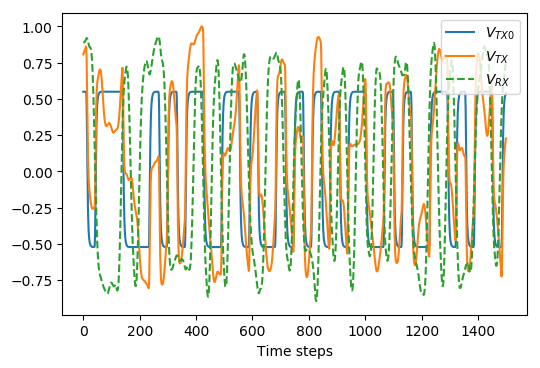} 
\par\end{centering}
\caption{Training data collected with the setup shown in \ref{fig:Measurement-setup-for}.\label{fig:Samples-of-trainingV}}
\end{figure}

\begin{figure}[h]
\begin{centering}
\includegraphics[scale=0.5]{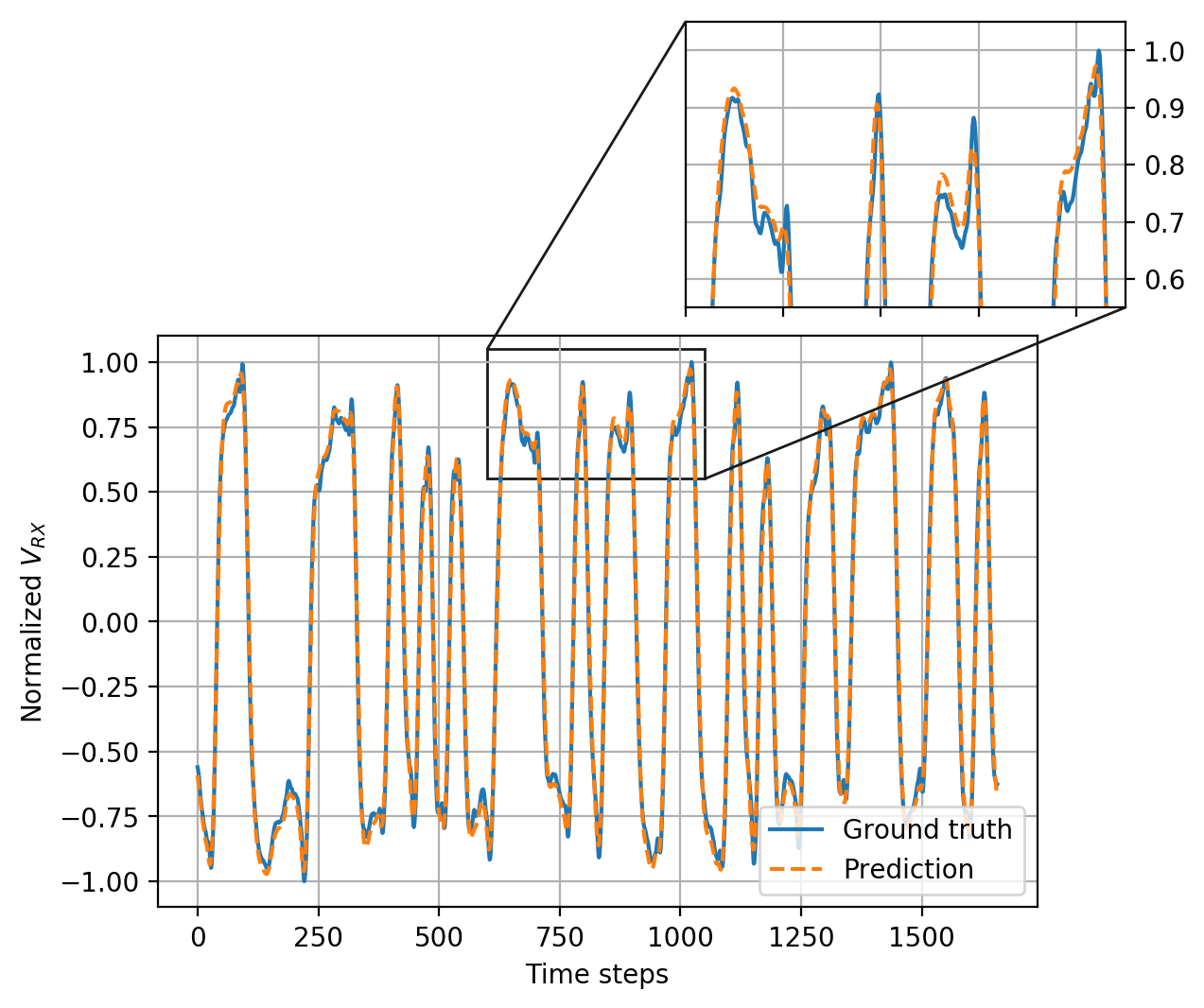} 
\par\end{centering}
\caption{Predicted voltage at the receiver $V_{RX}$ with a LSTM network. \label{fig:Prediction-of-RX}}
\end{figure}

The data is normalized and segmented into sequences of length $K$.
{Sample sequences after normalization of all the signals of interest are depicted}
in \ref{fig:Samples-of-trainingV}. This number $K$ represents the
memory dependency of the system. The larger the $K$ is, the longer
the memory the system keeps. A portion of the data ($10\%$) is reserved
for test. In this example, a stack of four LSTM cells of 20 hidden
units is used. The optimization method used is Adam with 0.3 dropout
regularization. Throughout our experiments, increasing $K$ not only
improves the convergence but also achieves higher accuracy. However,
once $K$ reaches the underlying memory length of the system under
learning, a further increase does not offer better convergence nor
higher accuracy. We use $K=10$ in the following numerical experiments.
The time steps for training is 11,000 and the model converges in about
48 epochs. Accurate predictions are achieved on unseen sequence as
shown in \ref{fig:Prediction-of-RX}.

\ref{fig:rnn_vs_lstm_k_4} and \ref{fig:rnn_vs_lstm_k_10} show the
comparison between LSTM network and vanilla RNN in terms of
their capability of handling the long-term memory. The same network
architecture is adopted in this comparison including the number of
layers, the layer width, and the regularization. It is shown that
when the memory is relatively short {with} $K=4$, the vanilla RNN cells
fails to capture the signal evolution whereas the LSTM network makes
pretty accurate predictions. When the memory is sufficiently
long, the vanilla RNN starts making comparably accurate predictions
as the LSTM network does, which is shown in \ref{fig:rnn_vs_lstm_k_10}.
From this comparison, it also reveals that training with Adam optimizer achieves
better performance than the SGD optimizer regardless of the memory
length.

\begin{center}
\begin{figure}[h]
\begin{centering}
\includegraphics[scale=0.4]{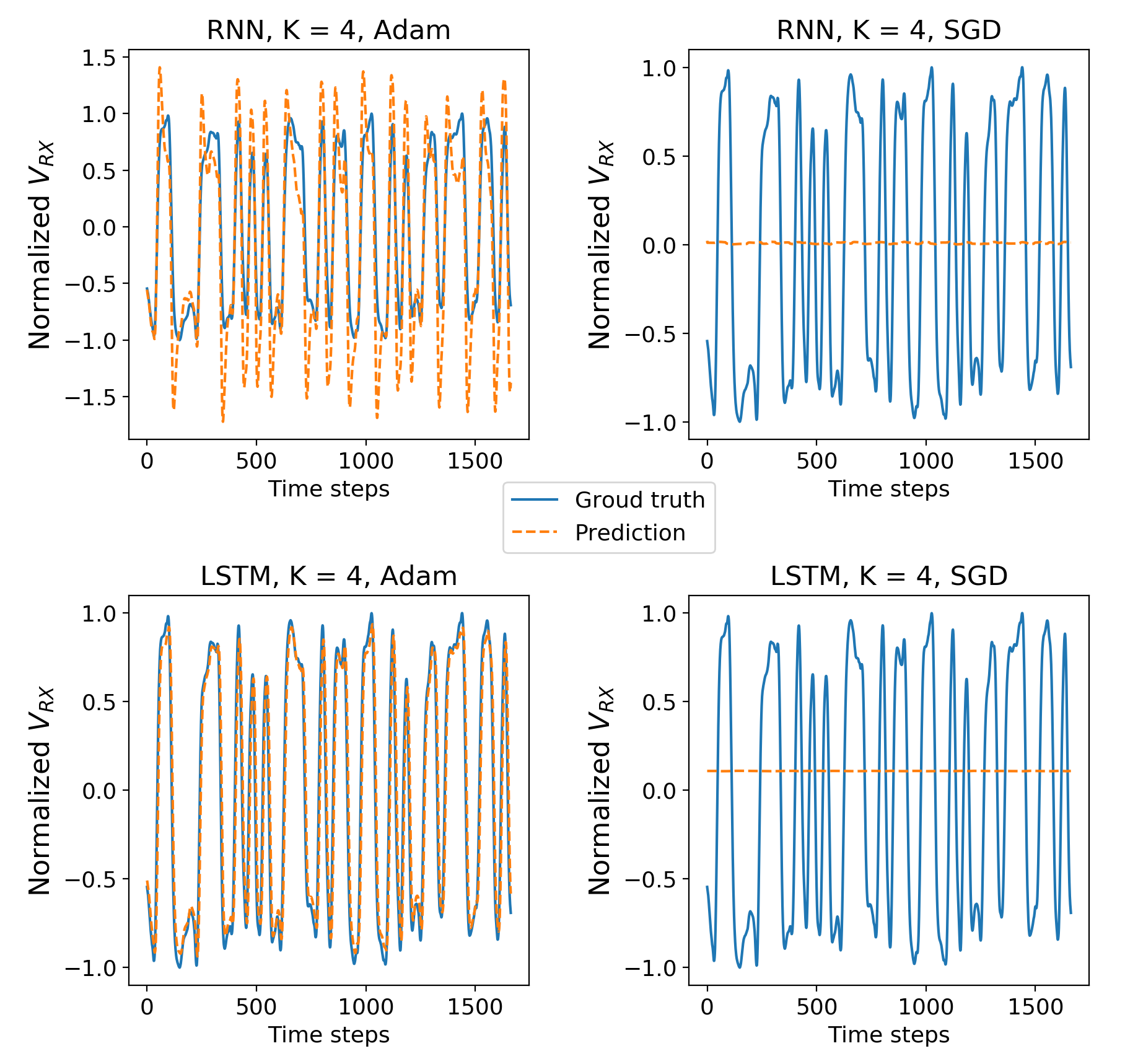} 
\par\end{centering}
\caption{Comparison between vanilla RNN and LSTM network in handling relative
short memory when the memory length $K$ is chosen as 4. \label{fig:rnn_vs_lstm_k_4}}
\end{figure}
\par\end{center}

\begin{figure}[ht]
\begin{centering}
\includegraphics[scale=0.4]{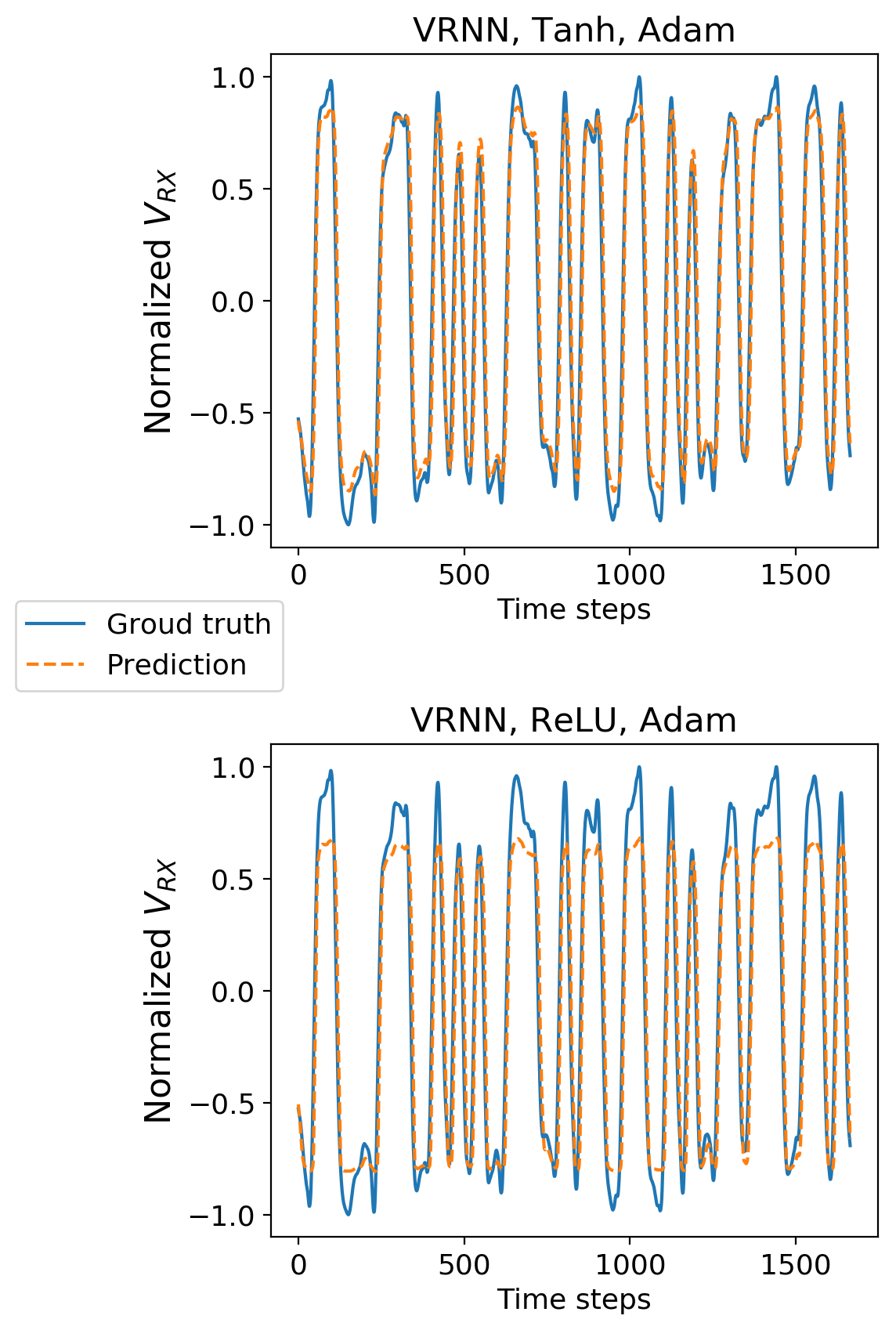} 
\par\end{centering}
\caption{The impact from different types of activation functions on the prediction
accuracy in a vanilla RNN. \label{fig:rnn_relu_vs_tanh}}
\end{figure}

It is worth mentioning that while using the LSTM and GRU networks,
one needs to pay particular attention upon the selection of activation
functions. For example, the gating signals $f_{t}$ and $i_{t}$ in
Equation (\ref{eq:lstm}) controls the percentage of the memory passing
through the gates, which ranges from 0 to 1. In this case, the activation
function associated with $f_{t}$ and $i_{t}$ has to be the sigmoid
function. Besides, the gating signal $g_{t}$ in Equation (\ref{eq:lstm})
allows both addition and subtraction operations between the input
and the forget gates and the hyperbolic tangent function {is appropriate}. As for a VRNN, the selection of activation
functions is only based on the nonlinearities. As shown in \ref{fig:rnn_relu_vs_tanh},
using the hyperbolic tangent function as the activation function in
a vanilla RNN achieves more accurate predictions than that with ReLU.

\begin{center}
\begin{figure}[ht]
\begin{centering}
\includegraphics[scale=0.4]{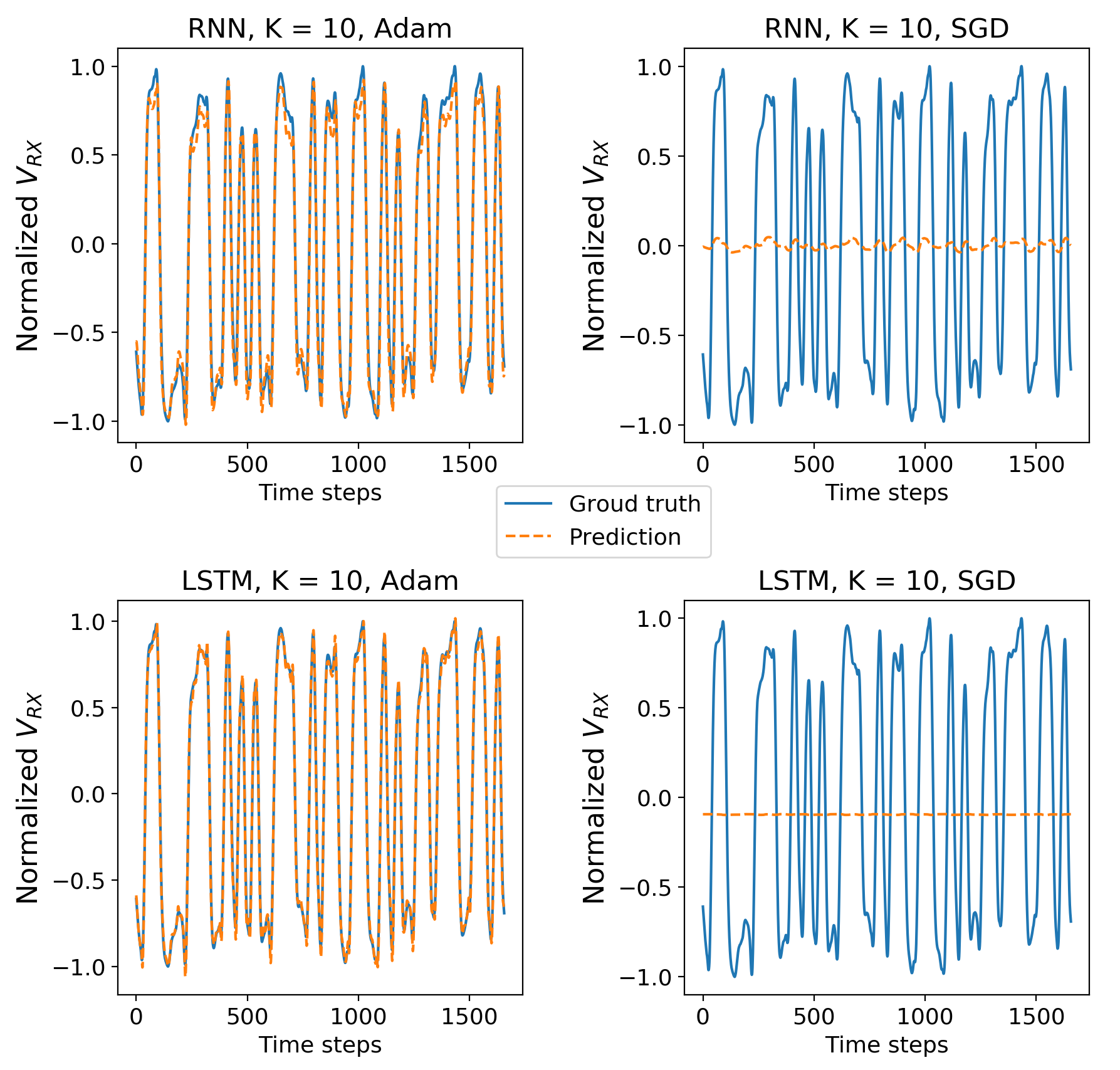} 
\par\end{centering}
\caption{Comparison between vanilla RNN and LSTM network in handling sufficiently
long memory when the memory length $K$ is chosen as 10.\label{fig:rnn_vs_lstm_k_10}}
\end{figure}
\par\end{center}

\begin{figure}[h]
\begin{centering}
\includegraphics[scale=0.4]{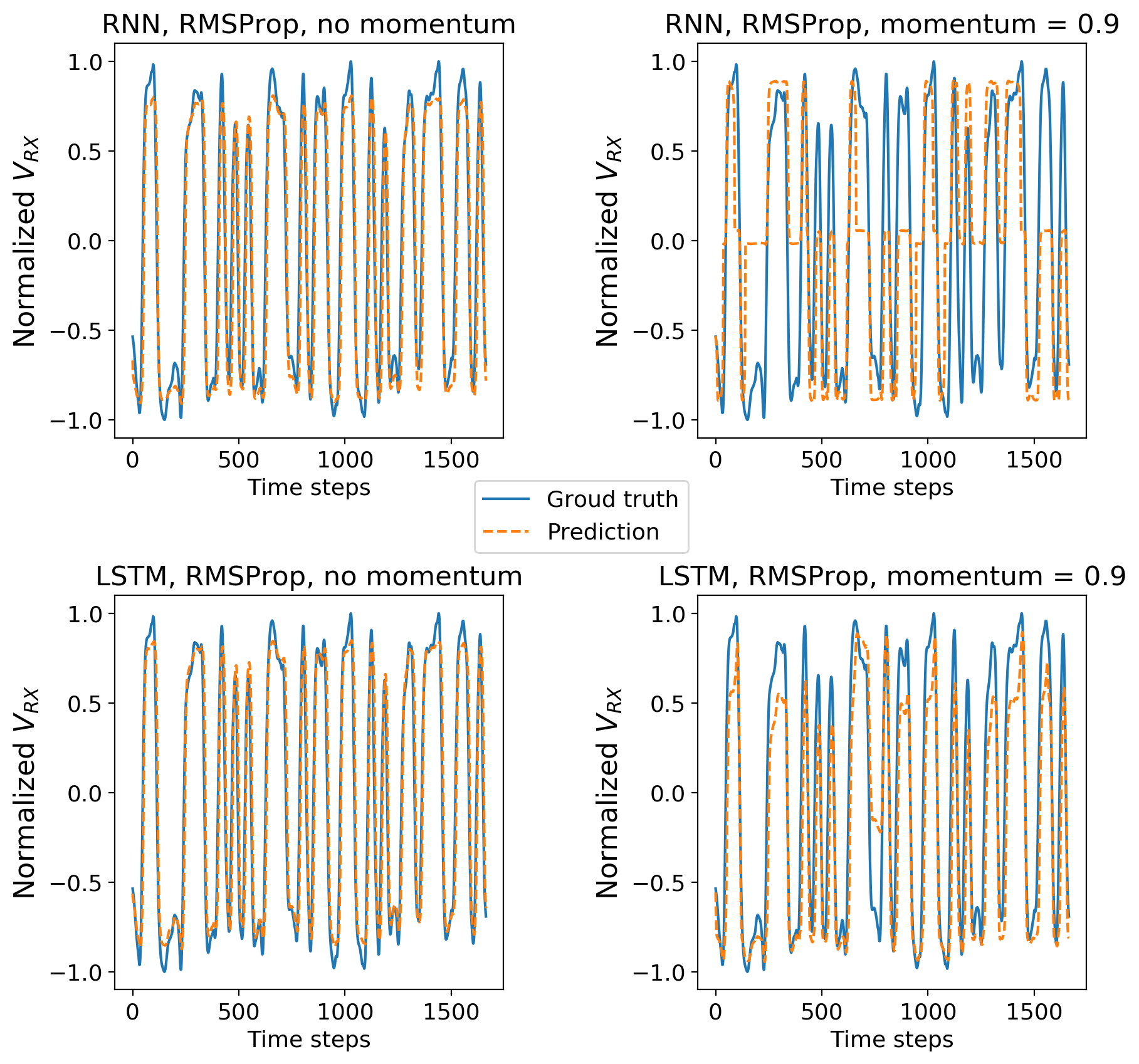} 
\par\end{centering}
\caption{Performance of the same architecture using different RNN cells, trained
by RMSProp when $K=5$\label{fig:rnn_rmsprop}}
\end{figure}

{In addition}, SGD optimizer {does not} work for the proposed RNN structure
{under the aforementioned settings for training}. Adding momentum for SGD does
not help the learning process either. However, Adam optimizer {achieves accurate predictions}. 
We also {investigate} RMSProp optimizer, which has been used for RNNs long before Adam
is invented, to train the same architecture {in terms of both} VRNN and
LSTM {network}. It {is found} that for short memory such as $K=4$,
{using} RMSProp optimizer {does not achieve convergence; as the
memory length $K$ goes beyond 5, RMSProp optimizer performs as
well as Adam}. The result shown in \ref{fig:rnn_rmsprop} confirms
that for $K=5$, networks trained by RMSProp make {accurate predictions} on the output waveform.
{However}, setting a high
momentum to deploy adaptive learning rate {degrades} the performance of
the network; {as can be seen in \ref{fig:rnn_rmsprop}, the prediction accuracy becomes worse with
momentum added in RMSProp.} 

\subsection{PAM4 channel simulation with ERNN}

\begin{figure}
	\begin{centering}
		\includegraphics[scale=0.4]{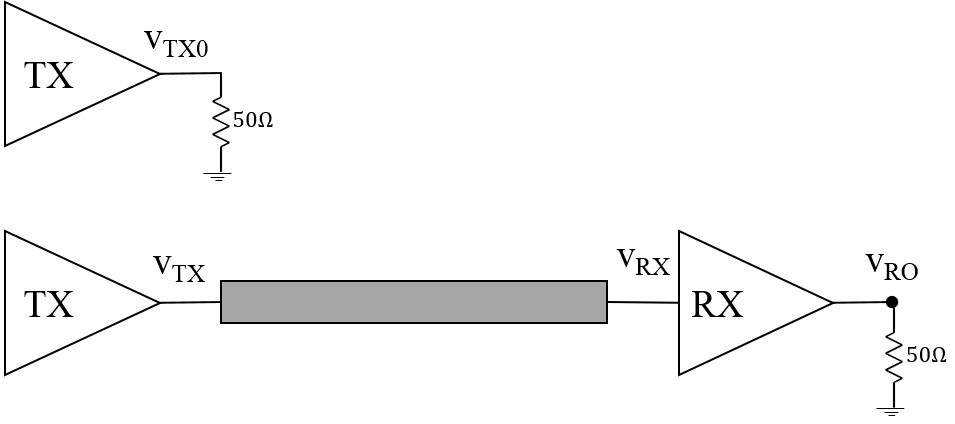} 
		\par\end{centering}
	\caption{Setup to obtain training data for PAM4 example.\label{fig:RX-learn}}
\end{figure}
The limitation of the {output-feedback RNN used in the PAM2 example is that
it strictly requires the output of the current time step before it can make predictions on one future time step,
which can be seen from Equation \eqref{eq:ar}.} The neural
network model, {while being used in this way}, cannot utilize batch inference.
In this {example}, it is shown that using a deeper and wider network,
an RNN-based model can be developed to utilize batch inference, 
{which can dramatically} reduce the run time for long transient simulation.

\begin{figure}
\begin{centering}
\includegraphics[scale=0.4]{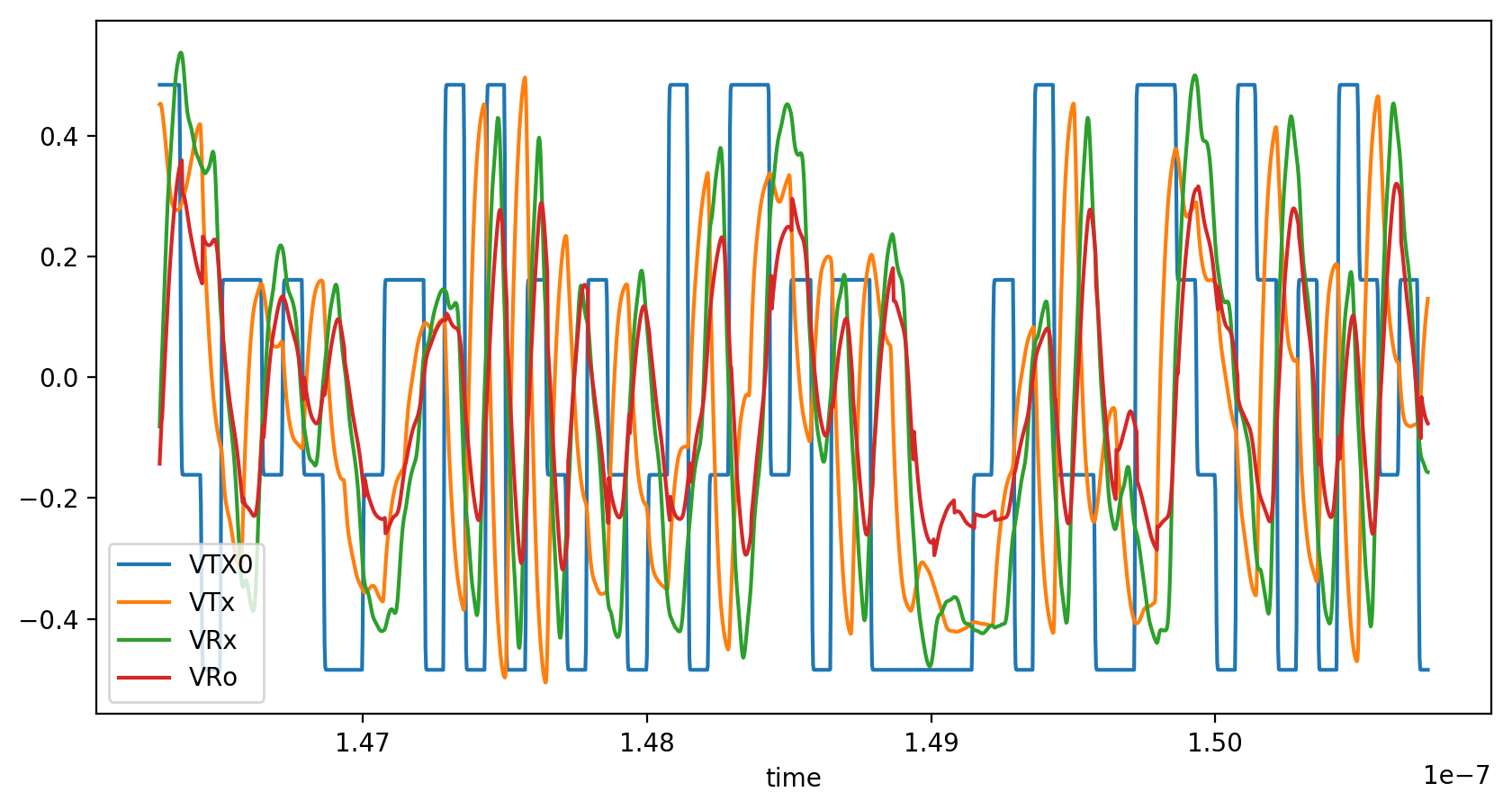} 
\par\end{centering}
\caption{Voltages used to train ERNN in PAM4 example.\label{fig:PAM4-train-data}}
\end{figure}

\begin{figure}
\begin{centering}
\includegraphics[scale=0.4]{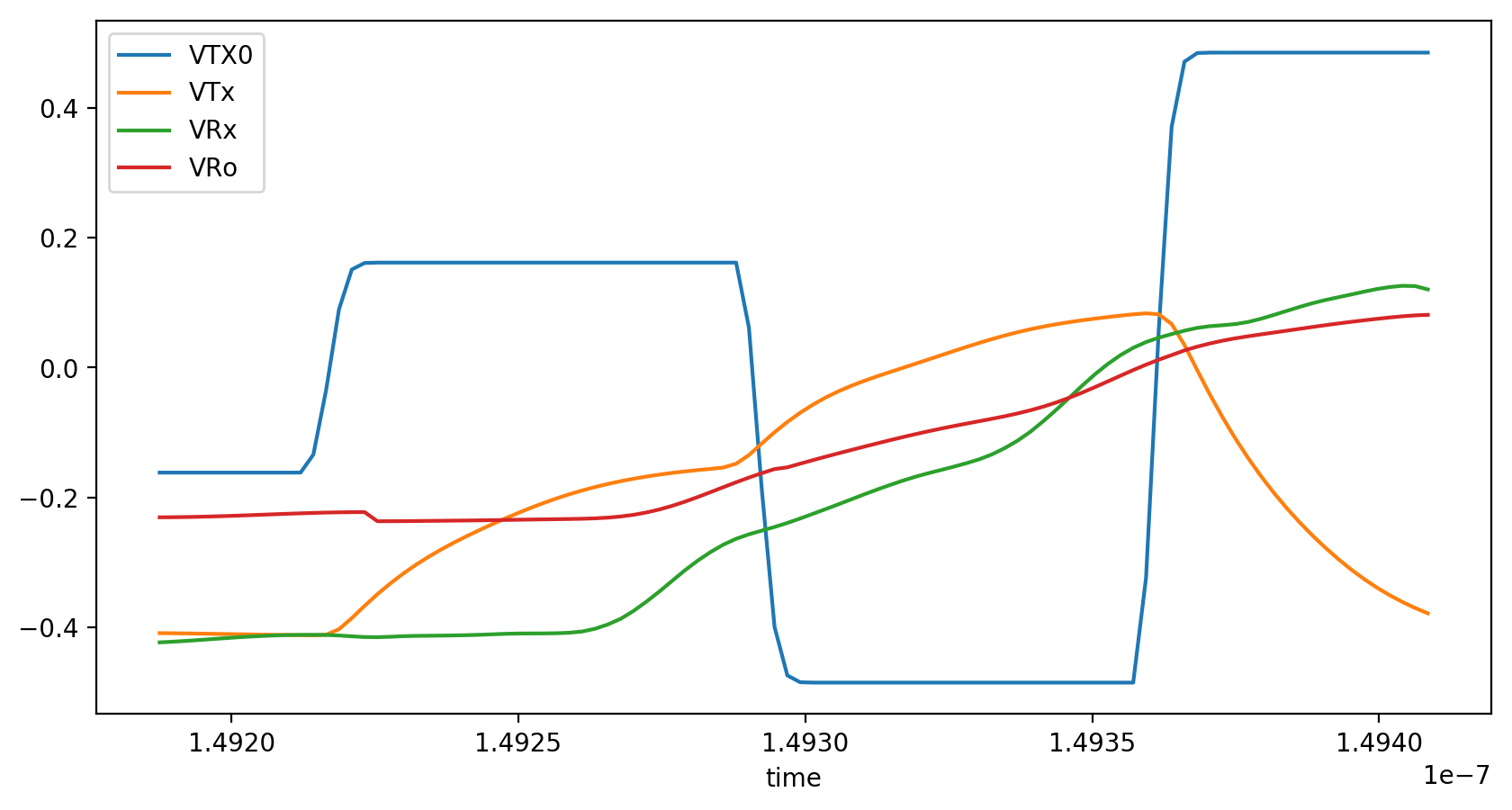} 
\par\end{centering}
\caption{A training sample by windowing the training sequence with $K=100$.\label{fig:PAM4-a_sample}}
\end{figure}

To prepare the training data, first, the transmitter output is measured
when it is opened, denoted {as} $V_{TX0}$. This signal is the Thevenin
source to the {combined} ``channel and receiver'' system {of interest}.
{When the transmitter is connected to the channel and the receiver,
the input to the channel from the transmitter $V_{TX}$ and the input to the receiver after the channel $V_{RX}$ are both collected for training purpose.}
{Besides}, the output voltage from the receiver $V_{RO}$ is also captured and included in the training set. 
The setup for data collection is shown in \ref{fig:RX-learn}. The data in this example comes from a PAM4 transceiver circuit transmitting data at 28 Gbps. An LSTM {network} is trained on about $10,000$ time points of time domain response of as shown in \ref{fig:PAM4-train-data}. 
{A training waveform sample is shown in \ref{fig:PAM4-a_sample}.}

\begin{figure*}
\begin{centering}
\subfloat[When $K=50$, trained in 100 epochs.\label{fig:pam4_K50_100epox}]{\begin{centering}
\includegraphics[scale=0.35]{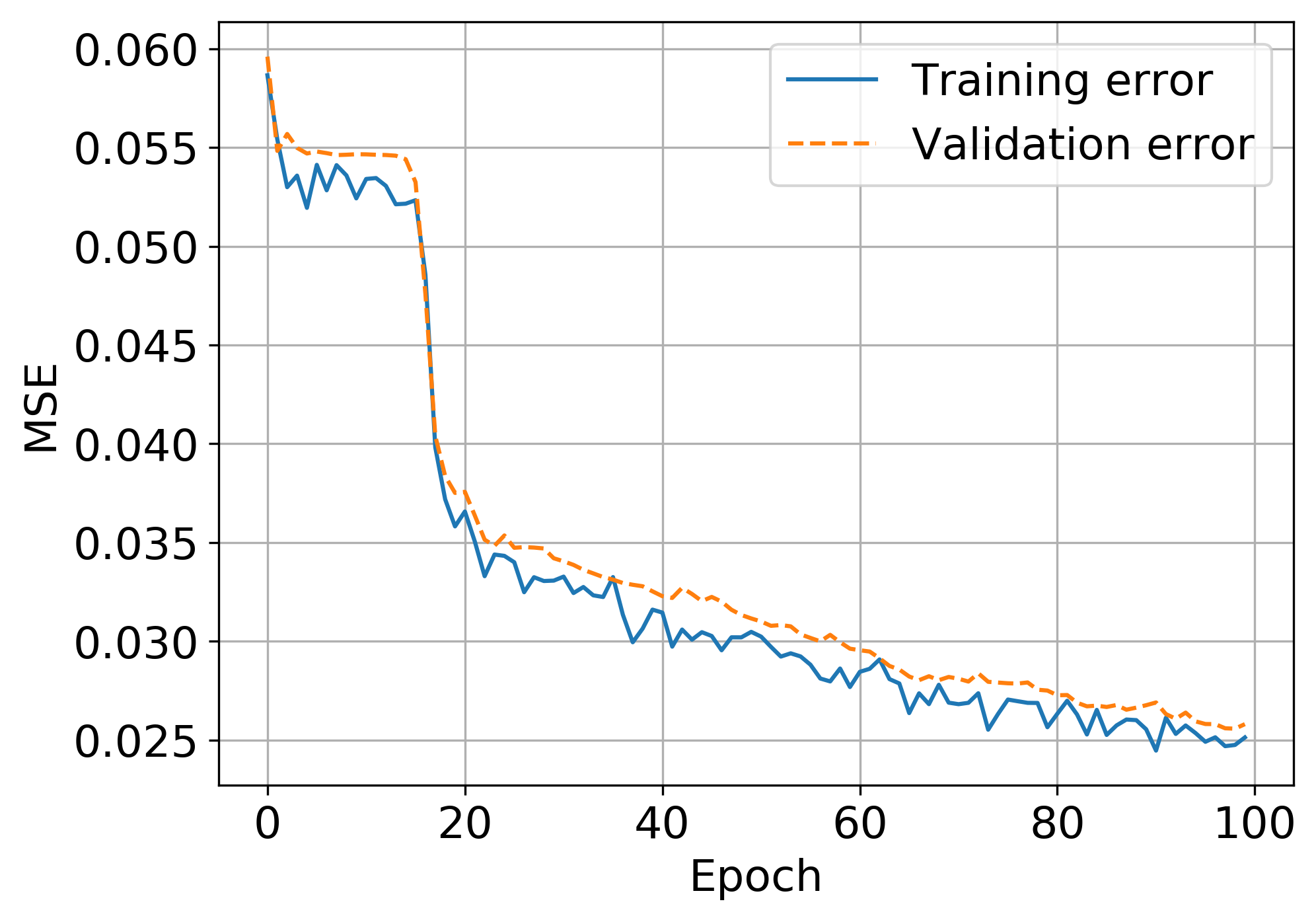}\hfill{}\includegraphics[scale=0.35]{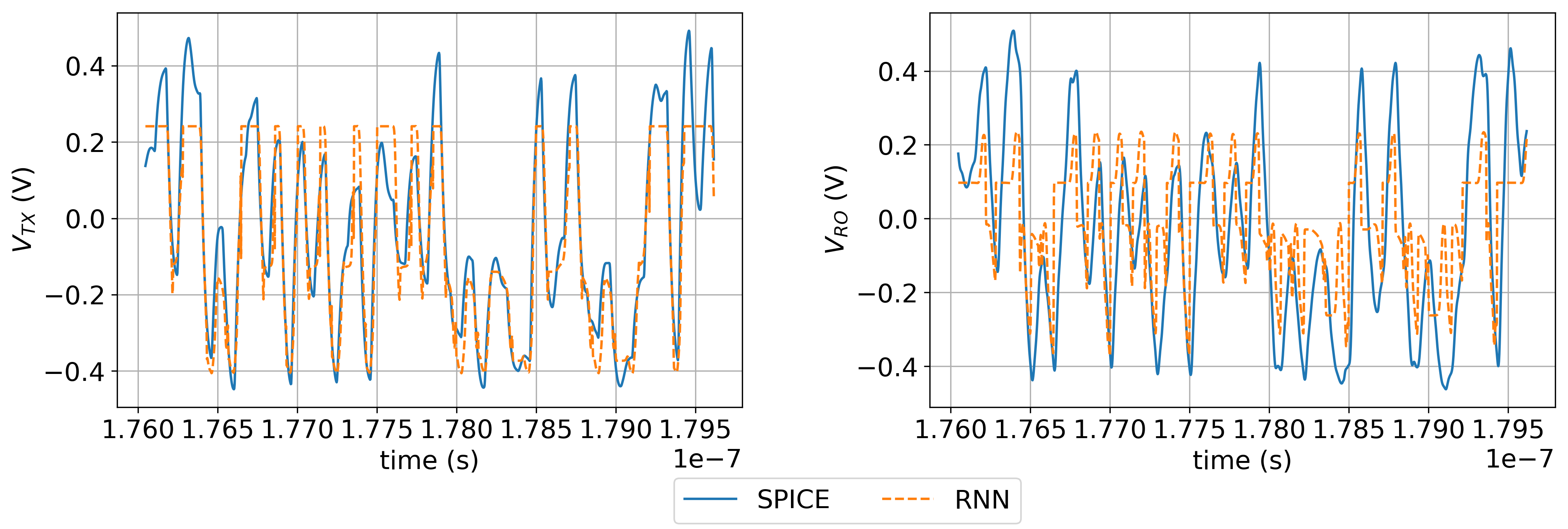} 
\par\end{centering}
}
\par\end{centering}
\begin{centering}
\subfloat[When $K=50$, trained in 1,000 epochs.\label{fig:pam4_K50_1000epox}]{\begin{centering}
\includegraphics[scale=0.35]{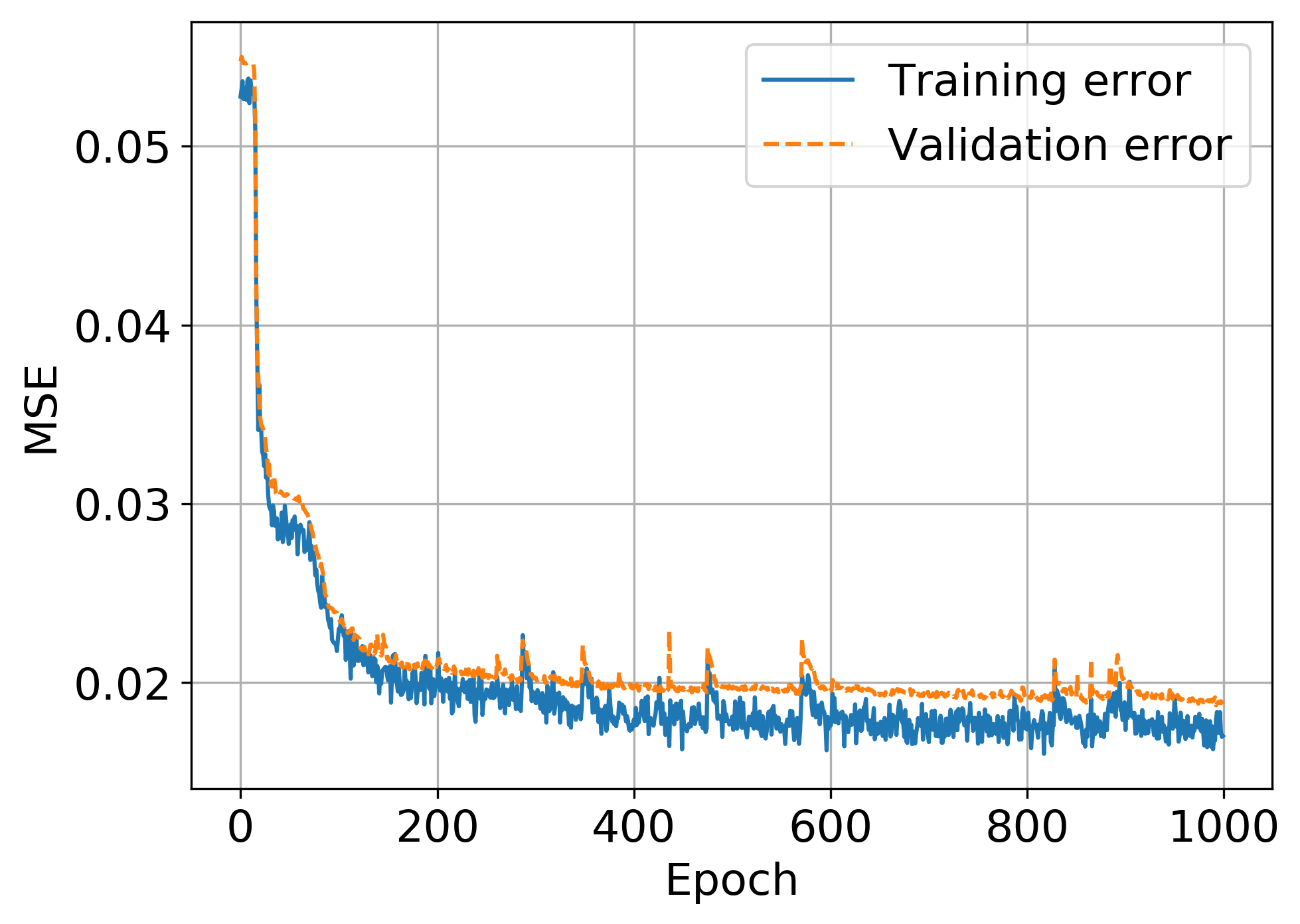}\hfill{}\includegraphics[scale=0.35]{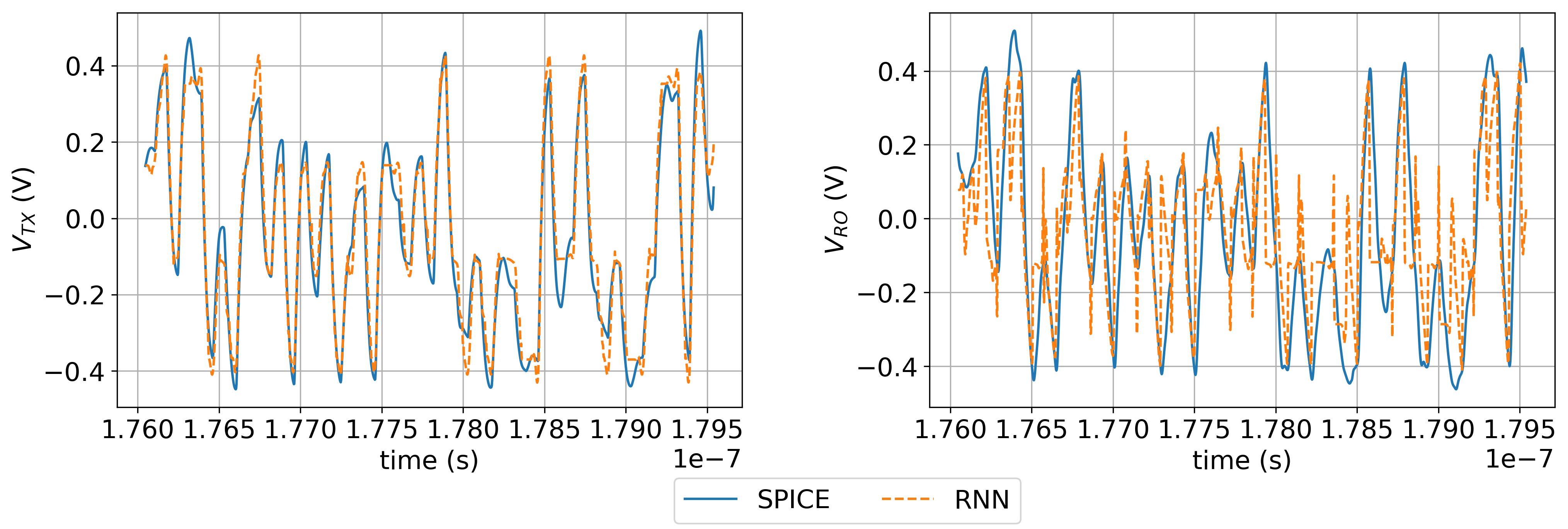} 
\par\end{centering}
}
\par\end{centering}
\begin{centering}
\subfloat[When $K=90$ trained in 1,000 epochs.\label{fig:pam4_K90_1000epox}]{\begin{centering}
\includegraphics[scale=0.35]{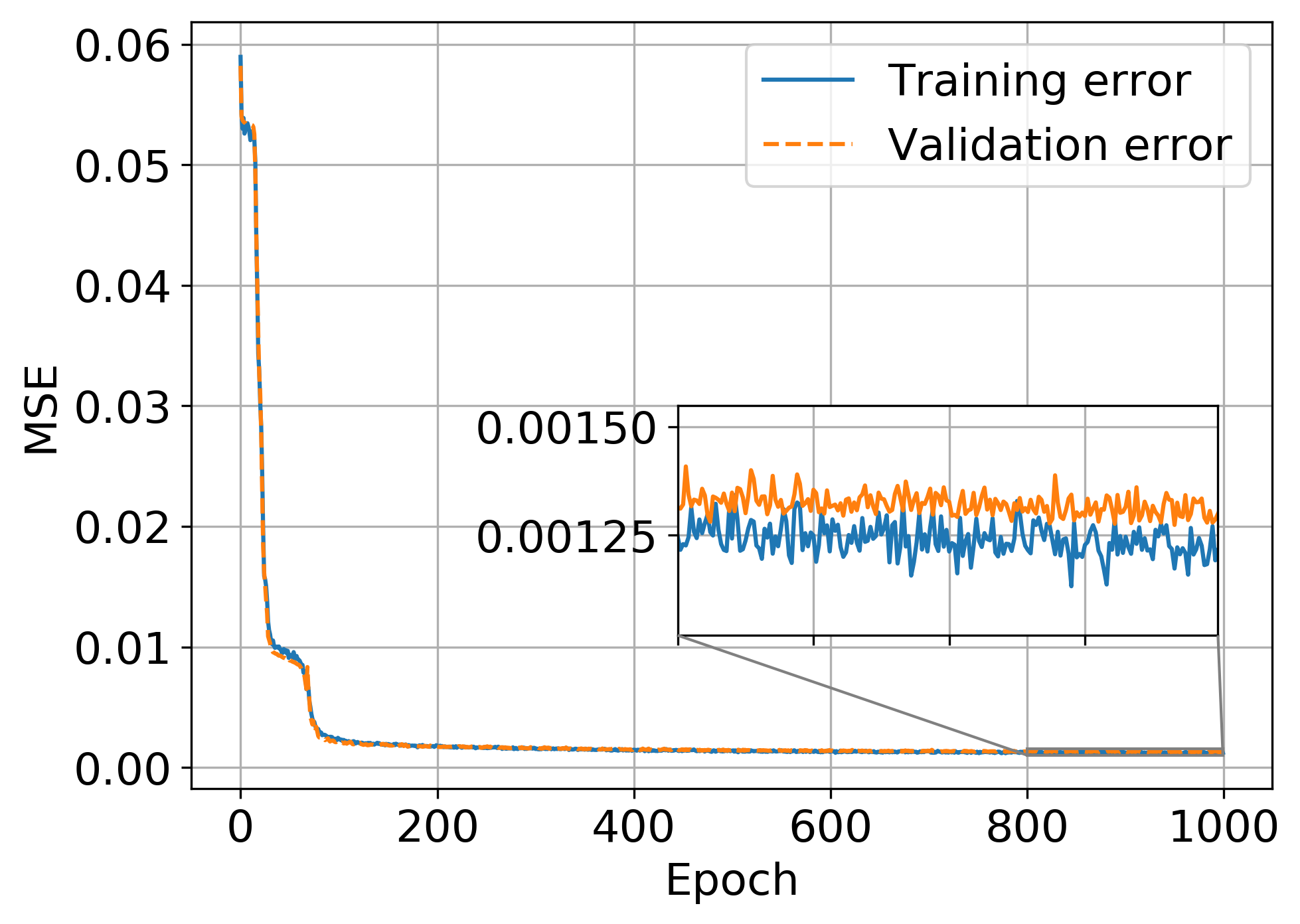}\hfill{}\includegraphics[scale=0.35]{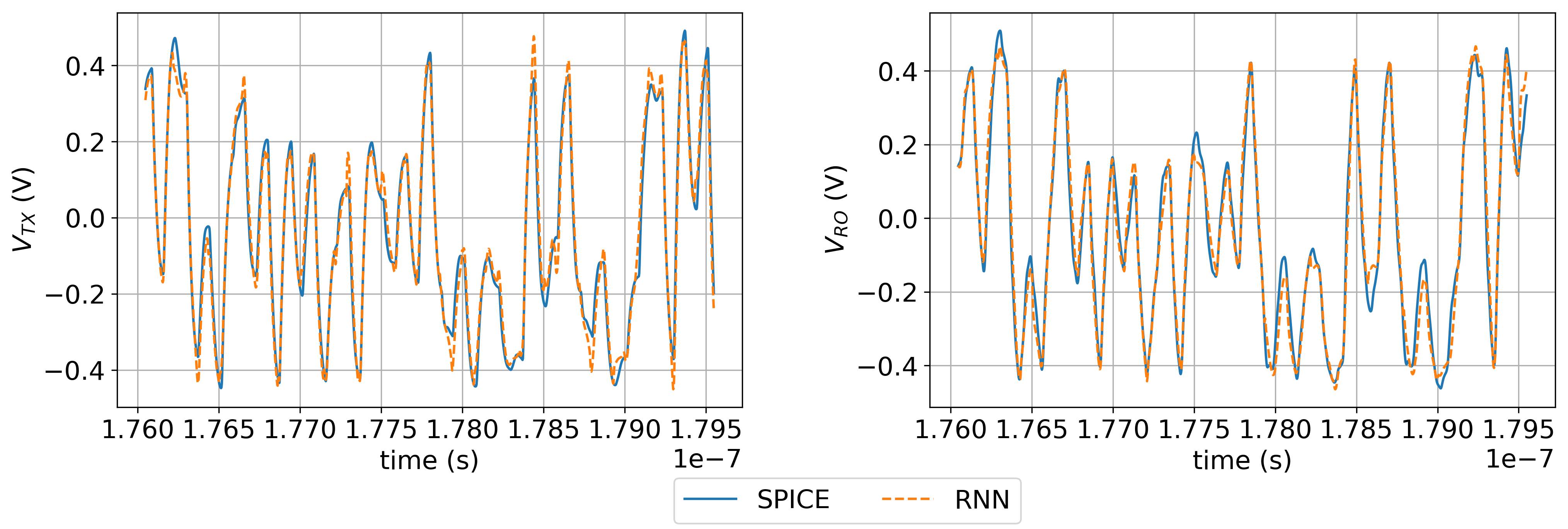} 
\par\end{centering}
}
\par\end{centering}
\begin{centering}
\subfloat[When $K=100$, trained in 1,000 epochs.\label{fig:pam4_K100_1000epox}]{\begin{centering}
\includegraphics[scale=0.35]{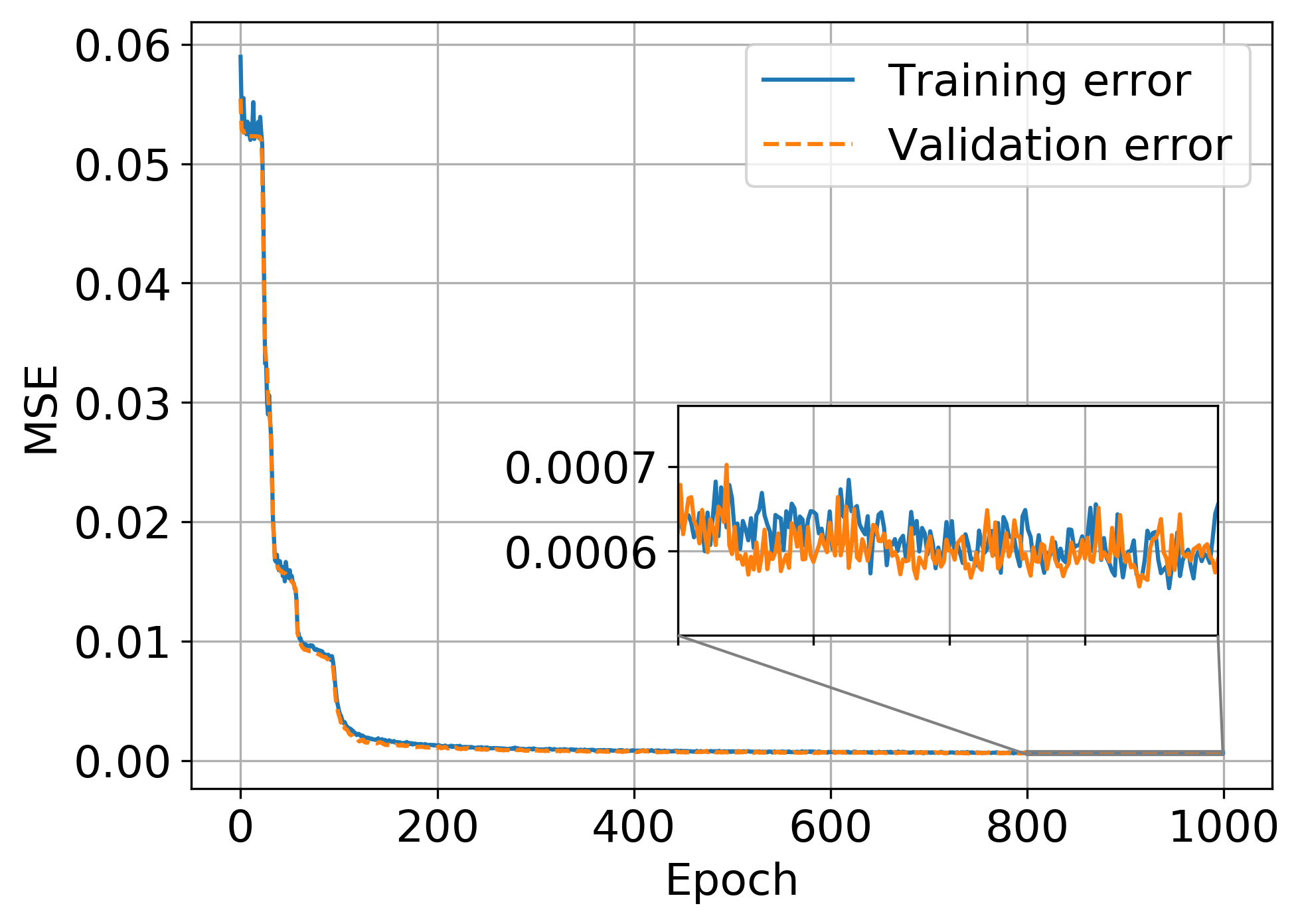}\hfill{}\includegraphics[scale=0.35]{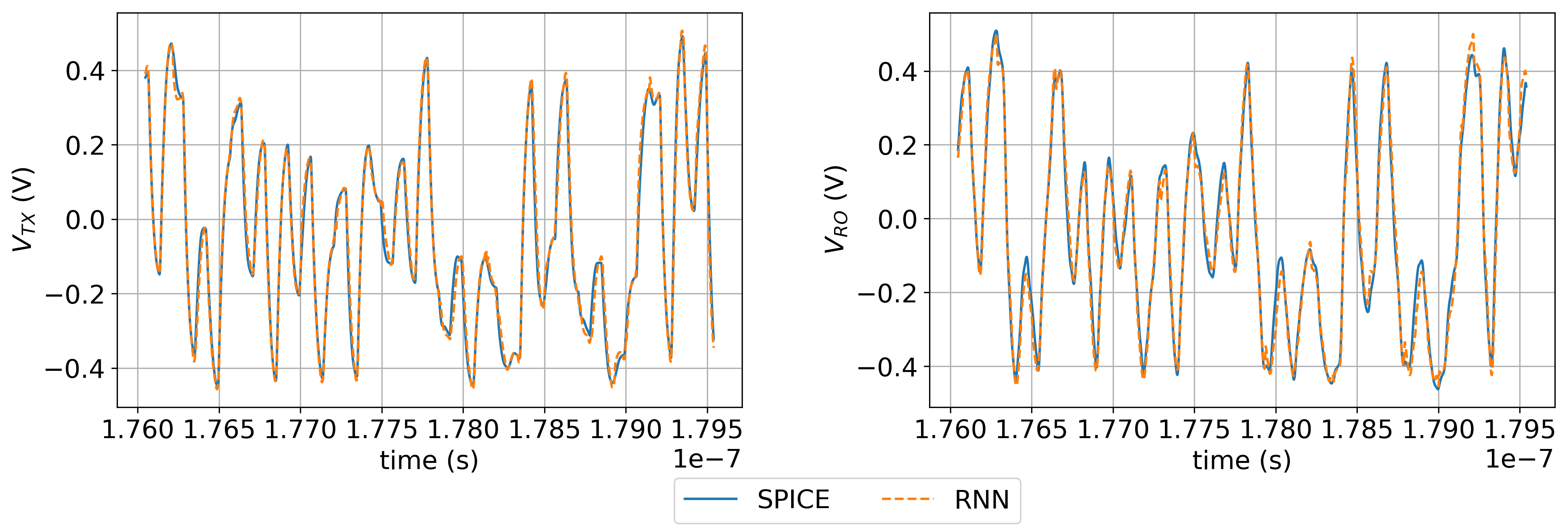} 
\par\end{centering}
}
\par\end{centering}
\caption{Training error (most left) and test performance of RNN model in PAM4
transceiver example.\label{fig:pam4_fullink}}
\end{figure*}

{We first investigate the impact from memory length on the training process.} The memory length {depends on} not only the nonlinearity of the transmitter and receiver but also the delay of the channel.
As for the training setup, Adam is used as the optimizer with initial learning rate
of 0.001 and dropout regularization is fixed {at} 0.3.
The LSTM network has six layers each with 30 hidden units.
The memory length $K$ is varied with everything else remaining the same.
\ref{fig:pam4_fullink} {demonstrates the training performance under various memory lengths with the
same network topology.
By showing the results at different epochs, \ref{fig:pam4_fullink} also reveals the fact that 
the learning ability of the LSTM network evolves as the training progresses.
For example, at the $100^{\textrm{th}}$ epoch, the LSTM network learned the switching pattern of the
waveforms; at the $1000^{\textrm{th}}$ epoch, the same network is able to make accurate predictions on $V_{TX}$ 
in terms of both the pattern and the amplitude.}

{With the memory length chosen as $K\!=\!50$ and at the $1000^{\textrm{th}}$ epoch, the predictions made with the LSTM network on $V_{RO}$
are less accurate than those on $V_{TX}$, as shown in \ref{fig:pam4_fullink}.
The reason that obtaining accurate predictions on $V_{RO}$ is more challenging than that for $V_{TX}$ is because the former requires
a much better knowledge of the delay imposed by the channel. 
It seems the memory length set by $K\!=\!50$ dose not provide adequate data on the channel delay.
After the memory length is increased to $K=90$, the predictions on $V_{RO}$ become much more accurate, as shown in \ref{fig:pam4_K90_1000epox}.
A further increase of the memory length to $K\!=\!100$ does not further improve the performance as shown in \ref{fig:pam4_K100_1000epox}.}
It is worth mentioning that the increase of memory length demands more computation resources.

\begin{figure}[ht]
\begin{centering}
\subfloat[From SPICE]{\begin{centering}
\includegraphics[scale=0.35]{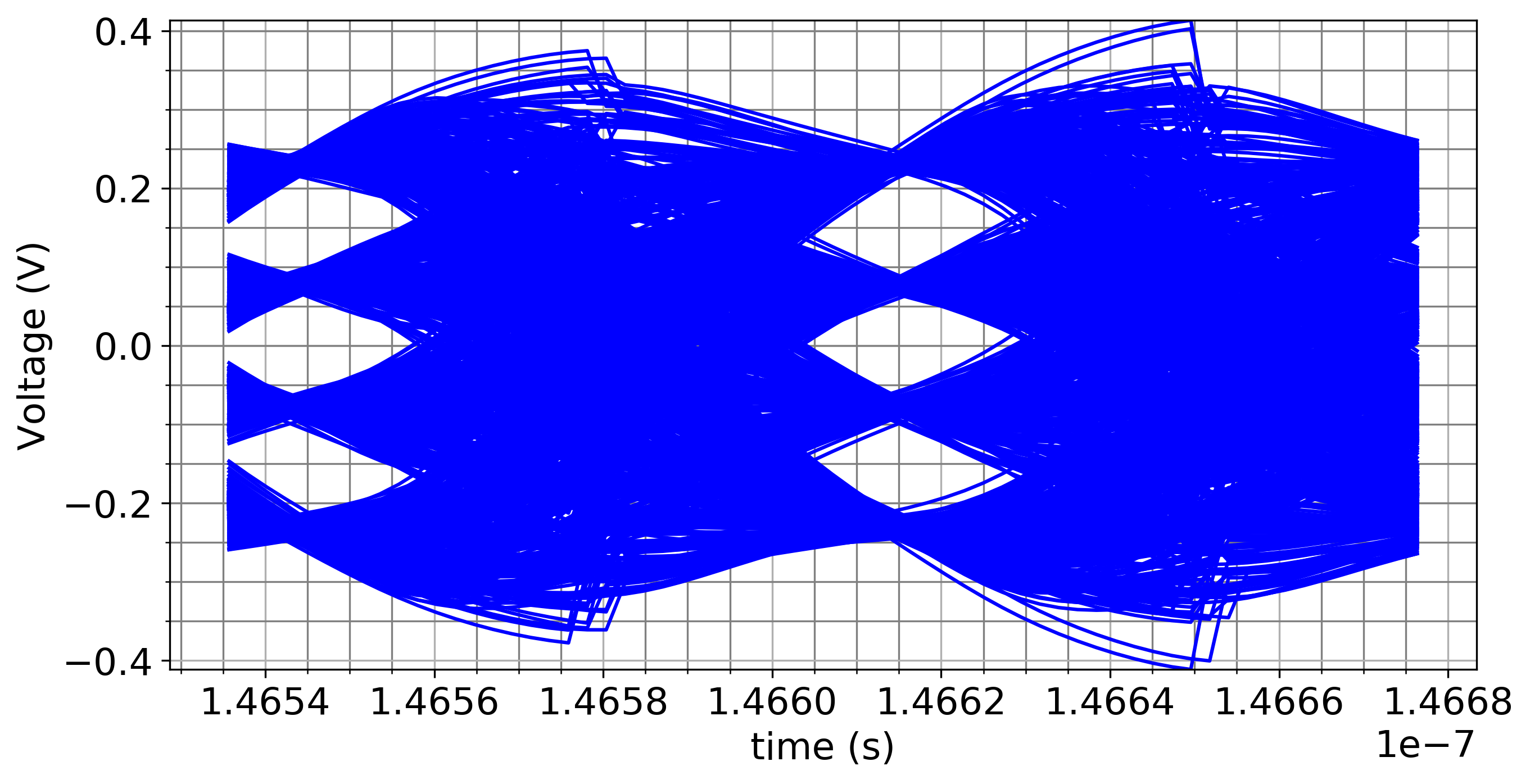} 
\par\end{centering}
}
\par\end{centering}
\begin{centering}
\subfloat[From RNN]{\begin{centering}
\includegraphics[scale=0.35]{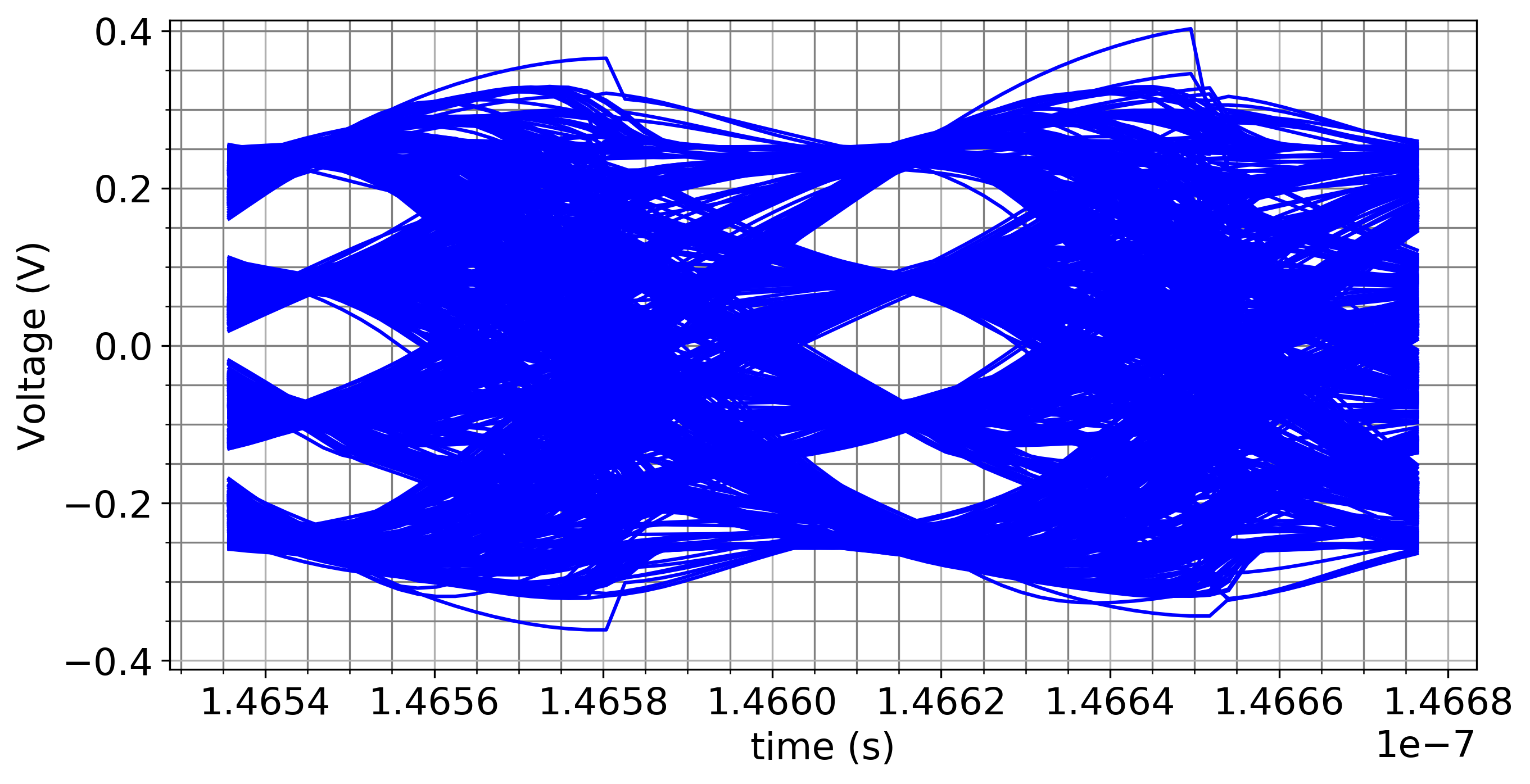} 
\par\end{centering}
}
\par\end{centering}
\caption{Eye diagram obtained in PAM4 transceiver example.\label{fig:pam4_eye}}
\end{figure}

{To further validate the model, we employ a much longer PRBS than
the training one and generate eye diagrams.} In \ref{fig:pam4_eye}, it shows a very good agreement between
the eye diagram generated from traditional SPICE-like simulation and the one from the proposed RNN-based model. \tlua{For example, both eye diagrams point out that the optimal sampling point is about 14.662 $\mu s$.}

\subsection{Accumulation of numerical error}

\begin{figure*}
	\begin{centering}
		\subfloat[Initially]{\begin{centering}
				\includegraphics[scale=0.1]{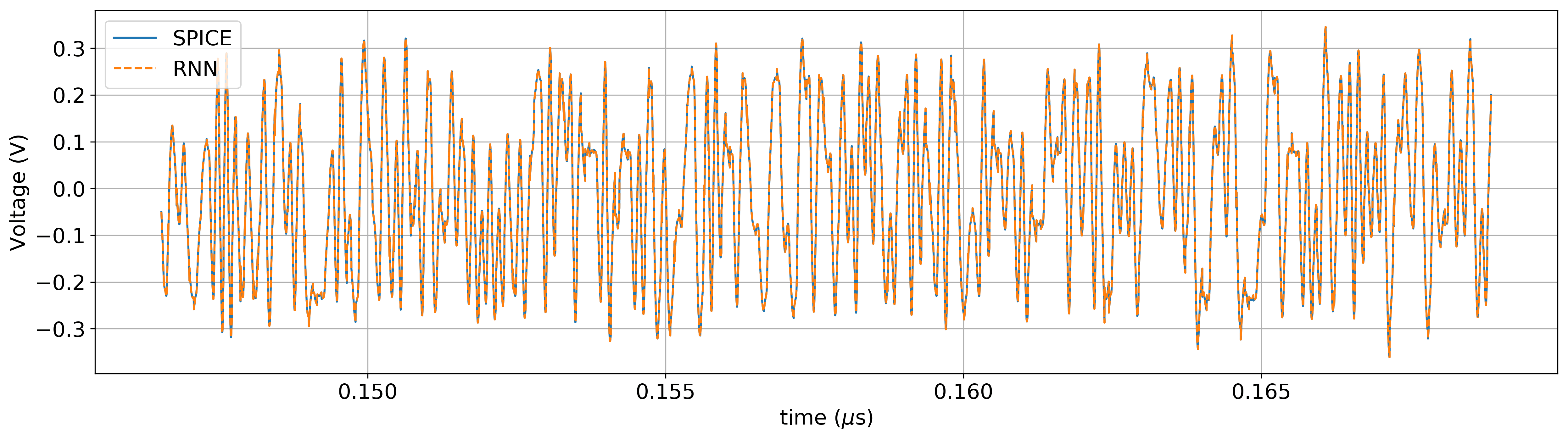}
				\par\end{centering}
		}
		\par\end{centering}
	\begin{centering}
		\subfloat[After tens of thousands of bits]{\begin{centering}
				\includegraphics[scale=0.1]{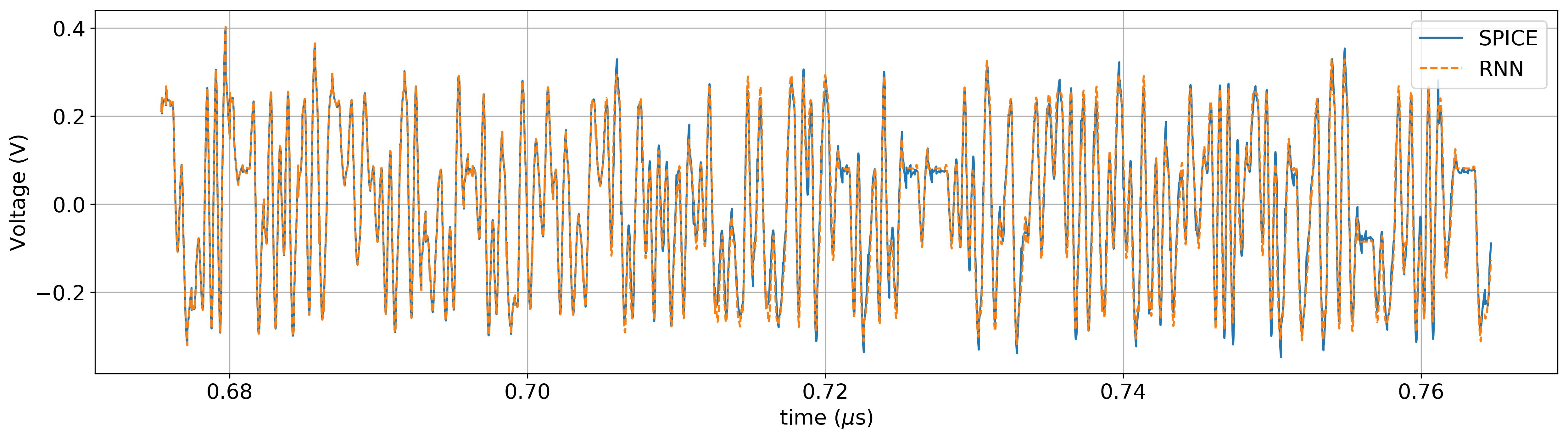}
				\par\end{centering}
		}
		\par\end{centering}
	\begin{centering}
		\subfloat[After hundreds of thousands of bits]{\begin{centering}
				\includegraphics[scale=0.1]{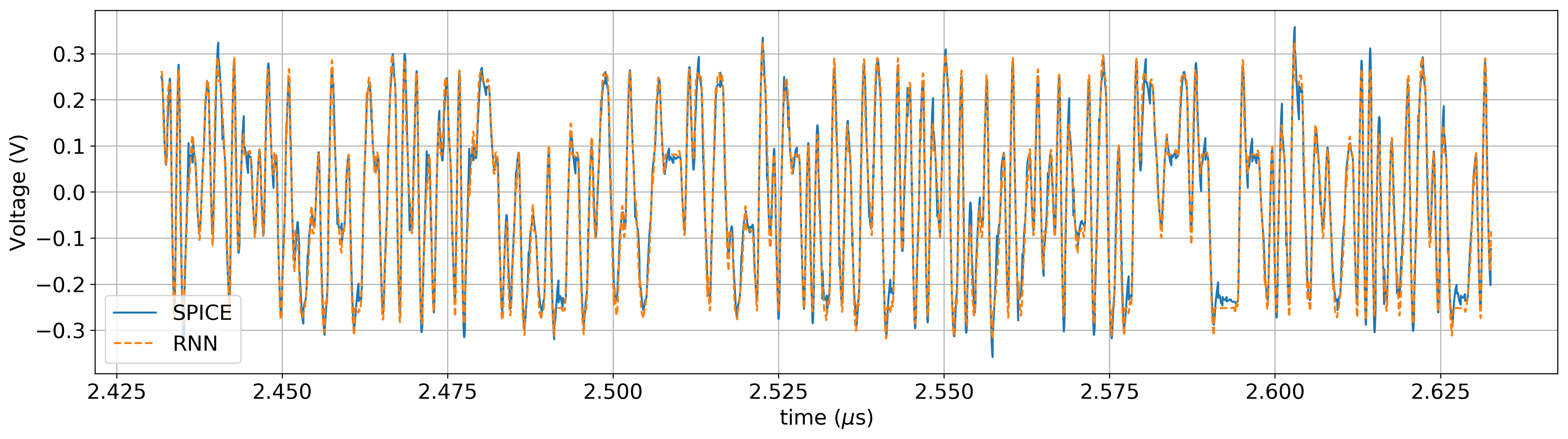}
				\par\end{centering}
		}
		\par\end{centering}
	\caption{Waveform comparison between SPICE simulation and RNN prediction on
		PAM4 example with another PRBS\label{fig:pam4_longPRBS}}.
\end{figure*}

{One limitation of the proposed method with RNN is the accumulation of numerical error. 
During the training process, TBPTT gives a noisy
gradient information to the optimizer, which translates to the numerical error in the solution.
This numerical error, though initially very small, gradually accumulates as the prediction goes on with the input sequence. 
A longer input sequence leads to a larger numerical error in the predicted results.
\ref{fig:pam4_longPRBS} shows the performance of the trained model in the previous section
on a very long PRBS. Initially, the predicted results from the RNN model agree well with those obtained from the circuit simulation. However, as the prediction progresses, the numerical error due to TBPTT accumulates and degrades the performance of the RNN model.
The accumulation of the numerical error is a well-known limitation of RNN trained by TBPTT, which at the same time leaves
room for improvement in the future work on the proposed method with advanced techniques for sequence
modeling such as attention mechanism \cite{RNN_attention}.}

\section{Conclusion and Future work\label{sec:Conclusion-and-Future}}

In this paper, we propose using RNN for transient high-speed link simulation. 
It shows that using RNN-based model for circuit simulation is promising in terms of both the accuracy and the capability of improving computation efficiency.
Through the proposed approach, an RNN model is trained and
validated on a relatively short sequence generated from a circuit
simulator. After the training completes, the RNN can be used to make
predictions on the remaining sequence to generate an eye
diagram. 
Using RNN model significantly enhances the computation efficiency because the transient waveforms are produced
through inference, which saves iterations in solving nonlinear
systems required by a circuit simulator.
An RNN differs from a FNN by the fact that its parameters are shared across time. 
This is an advantage of an RNN because the number of tunable parameters is significantly reduced comparing to a FNN.
However, it also becomes a challenge to train an RNN as the regular back-prop does not work anymore; instead, back-prop through time must be employed. 
Two topologies of the RNN, namely, ERNN and NARX-RNN, are investigated and compared in terms of the performance in high-speed link simulation. 
Through examples, it is demonstrated that ERNN without output feed-back is preferable in high-speed link simulation owing to its capability of batch learning and inference. 
It is also found out that LSTM network outperforms the vanilla RNN in terms of accuracy.
We also investigates the impacts of training schemes and tunable parameters on both the accuracy and the generalization capability of an RNN model through examples.

Understanding the memory length of the data for training is important in achieving a balance
between the computational cost and the accuracy of an RNN model. This remains
an open problem in causal inference domain and the selection of a sufficient memory length to train an RNN model heavily relies on prior experience and substantial domain knowledge.
One future work is to divide the full-channel modeling task into blocks where each block is represented by a standalone RNN model. 
Individual blocks can thus be swapped in and out to combine with different channel designs without retraining the RNN. 
Another future work is to take the equalization settings as 
inputs such that the RNN models can completely replace the transceivers circuit
models.

\section*{Acknowledgment}

This material is based upon work supported by the National Science
Foundation under Grant No. CNS 16-24810, the U.S Army Small Business
Innovation Research (SBIR) Program office and the U.S. Army Research
Office under Contract No.W911NF-16-C-0125 and by Zhejiang University
under grant ZJU Research 083650.

\bibliographystyle{IEEEtran}
\bibliography{rnn_epeps18}

\begin{thebibliography}{10}
\providecommand{\url}[1]{#1}
\csname url@samestyle\endcsname
\providecommand{\newblock}{\relax}
\providecommand{\bibinfo}[2]{#2}
\providecommand{\BIBentrySTDinterwordspacing}{\spaceskip=0pt\relax}
\providecommand{\BIBentryALTinterwordstretchfactor}{4}
\providecommand{\BIBentryALTinterwordspacing}{\spaceskip=\fontdimen2\font plus
\BIBentryALTinterwordstretchfactor\fontdimen3\font minus
  \fontdimen4\font\relax}
\providecommand{\BIBforeignlanguage}[2]{{%
\expandafter\ifx\csname l@#1\endcsname\relax
\typeout{** WARNING: IEEEtran.bst: No hyphenation pattern has been}%
\typeout{** loaded for the language `#1'. Using the pattern for}%
\typeout{** the default language instead.}%
\else
\language=\csname l@#1\endcsname
\fi
#2}}
\providecommand{\BIBdecl}{\relax}
\BIBdecl

\bibitem{SPICE_DeHon}
N.~Kapre and A.~DeHon, ``Parallelizing sparse matrix solve for {SPICE} circuit
  simulation using {FPGA}s,'' in \emph{2009 International Conference on
  Field-Programmable Technology}, Dec 2009, pp. 190--198.

\bibitem{GPU_SPICE_Yang}
X.~Chen, L.~Ren, Y.~Wang, and H.~Yang, ``{GPU-Accelerated Sparse LU
  Factorization for Circuit Simulation with Performance Modeling},'' \emph{IEEE
  Transactions on Parallel and Distributed Systems}, vol.~26, no.~3, pp.
  786--795, March 2015.

\bibitem{EyeGen_RBWu}
W.~D. Guo, J.~H. Lin, C.~M. Lin, T.~W. Huang, and R.~B. Wu, ``Fast methodology
  for determining eye diagram characteristics of lossy transmission lines,''
  \emph{IEEE Transactions on Advanced Packaging}, vol.~32, no.~1, pp. 175--183,
  Feb 2009.

\bibitem{RNN_app_uber}
\BIBentryALTinterwordspacing
L.~Zhu and N.~Laptev, ``Deep and confident prediction for time series at
  {Uber},'' in \emph{2017 IEEE International Conference on Data Mining
  Workshops (ICDMW)}, vol.~00, Nov. 2018, pp. 103--110. [Online]. Available:
  \url{doi.ieeecomputersociety.org/10.1109/ICDMW.2017.19}
\BIBentrySTDinterwordspacing

\bibitem{RNN_app_seq2seq2}
\BIBentryALTinterwordspacing
K.~Cho, B.~van Merrienboer, {\c{C}}.~G{\"{u}}l{\c{c}}ehre, F.~Bougares,
  H.~Schwenk, and Y.~Bengio, ``Learning phrase representations using {RNN}
  encoder-decoder for statistical machine translation,'' \emph{CoRR}, vol.
  abs/1406.1078, 2014. [Online]. Available:
  \url{http://arxiv.org/abs/1406.1078}
\BIBentrySTDinterwordspacing

\bibitem{RNN_app_seq2seq}
\BIBentryALTinterwordspacing
I.~Sutskever, O.~Vinyals, and Q.~V. Le, ``Sequence to sequence learning with
  neural networks,'' in \emph{Proceedings of the 27th International Conference
  on Neural Information Processing Systems - Volume 2}, ser. NIPS'14.\hskip 1em
  plus 0.5em minus 0.4em\relax Cambridge, MA, USA: MIT Press, 2014, pp.
  3104--3112. [Online]. Available:
  \url{http://dl.acm.org/citation.cfm?id=2969033.2969173}
\BIBentrySTDinterwordspacing

\bibitem{RNN_app_imgreg}
\BIBentryALTinterwordspacing
J.~Ba, V.~Mnih, and K.~Kavukcuoglu, ``Multiple object recognition with visual
  attention,'' \emph{CoRR}, vol. abs/1412.7755, 2014. [Online]. Available:
  \url{http://arxiv.org/abs/1412.7755}
\BIBentrySTDinterwordspacing

\bibitem{RNN_app_seqgen}
\BIBentryALTinterwordspacing
A.~Graves, ``Generating sequences with recurrent neural networks,''
  \emph{CoRR}, vol. abs/1308.0850, 2013. [Online]. Available:
  \url{http://arxiv.org/abs/1308.0850}
\BIBentrySTDinterwordspacing

\bibitem{RNN_narx_rfckt_review}
W.~Liu, W.~Na, L.~Zhu, and Q.~J. Zhang, ``A review of neural network based
  techniques for nonlinear microwave device modeling,'' in \emph{2016 IEEE
  MTT-S International Conference on Numerical Electromagnetic and Multiphysics
  Modeling and Optimization (NEMO)}, July 2016, pp. 1--2.

\bibitem{RNN_ERNN1}
A.~Beg, P.~W.~C. Prasad, M.~M. Arshad, and K.~Hasnain, ``Using recurrent neural
  networks for circuit complexity modeling,'' in \emph{2006 IEEE International
  Multitopic Conference}, Dec 2006, pp. 194--197.

\bibitem{RNN_ERNN2}
W.-T. Hsieh, C.-C. Shiue, and C.~N.~J. Liu, ``A novel approach for high-level
  power modeling of sequential circuits using recurrent neural networks,'' in
  \emph{2005 IEEE International Symposium on Circuits and Systems}, May 2005,
  pp. 3591--3594 Vol. 4.

\bibitem{RNN_ERNNzaichen}
Z.~Chen, M.~Raginsky, and E.~Rosenbaum, ``Verilog-{A} compatible recurrent
  neural network model for transient circuit simulation,'' in \emph{2017 IEEE
  26th Conference on Electrical Performance of Electronic Packaging and Systems
  (EPEPS)}, Oct 2017, pp. 1--3.

\bibitem{Eye_DNN_Lu}
T.~Lu, J.~Sun, K.~Wu, and Z.~Yang, ``{High-Speed Channel Modeling With Machine
  Learning Methods for Signal Integrity Analysis},'' \emph{IEEE Transactions on
  Electromagnetic Compatibility}, pp. 1--8, 2018.

\bibitem{Eye_S_FNN_Nikita}
N.~Ambasana, G.~Anand, D.~Gope, and B.~Mutnury, ``{S-Parameter and Frequency
  Identification Method for ANN-Based Eye-Height/Width Prediction},''
  \emph{IEEE Transactions on Components, Packaging and Manufacturing
  Technology}, vol.~7, no.~5, pp. 698--709, May 2017.

\bibitem{Eye_Spar_FNN_Goh}
C.~Goay, P.~Goh, N.~Ahmad, and M.~Ain, ``{Eye-height/width prediction using
  artificial neural networks from S-Parameters with vector fitting},''
  \emph{Journal of Engineering Science and Technology}, vol.~13, no.~3, pp.
  625--639, Mar. 2018.

\bibitem{NN_geom_2_S_ustrip}
G.~Xue-lian, C.~Zhen-nan, F.~Nan, Z.~Xiao-yu, and H.~Jian-hong, ``{An
  artificial neural network model for S-parameter of microstrip line},'' in
  \emph{2013 Asia-Pacific Symposium on Electromagnetic Compatibility (APEMC)},
  May 2013, pp. 1--4.

\bibitem{NN_geom_2_S_via}
X.~Zhang, Y.~Cao, and Q.~J. Zhang, ``A combined transfer function and neural
  network method for modeling via in multilayer circuits,'' in \emph{2008 51st
  Midwest Symposium on Circuits and Systems}, Aug 2008, pp. 73--76.

\bibitem{SFNN}
T.~Nguyen and J.~Schutt-Aine, ``{A Pseudo-supervised Machine Learning Approach
  to Broadband LTI Macro-Modeling},'' in \emph{2018 Joint Electromagnetic
  Compatibility (EMC) and Asia-Pacific Electromagnetic Compatibility (APEMC)
  Symposium}, 2018.

\bibitem{NN_poleres}
F.~Feng, C.~Zhang, J.~Ma, and Q.~J. Zhang, ``{Parametric Modeling of EM
  Behavior of Microwave Components Using Combined Neural Networks and
  Pole-Residue-Based Transfer Functions},'' \emph{IEEE Transactions on
  Microwave Theory and Techniques}, vol.~64, no.~1, pp. 60--77, Jan. 2016.

\bibitem{DL_ReLU}
\BIBentryALTinterwordspacing
X.~Glorot, A.~Bordes, and Y.~Bengio, ``Deep sparse rectifier neural networks,''
  in \emph{Proceedings of the Fourteenth International Conference on Artificial
  Intelligence and Statistics}, ser. Proceedings of Machine Learning Research,
  G.~Gordon, D.~Dunson, and M.~Dudík, Eds., vol.~15.\hskip 1em plus 0.5em
  minus 0.4em\relax Fort Lauderdale, FL, USA: PMLR, 11--13 Apr 2011, pp.
  315--323. [Online]. Available:
  \url{http://proceedings.mlr.press/v15/glorot11a.html}
\BIBentrySTDinterwordspacing

\bibitem{DL_Adam}
\BIBentryALTinterwordspacing
D.~P. Kingma and J.~Ba, ``Adam: {A} method for stochastic optimization,''
  \emph{CoRR}, vol. abs/1412.6980, 2014. [Online]. Available:
  \url{http://arxiv.org/abs/1412.6980}
\BIBentrySTDinterwordspacing

\bibitem{DL_SGD}
H.~Robbins and S.~Monro, ``{A Stochastic Approximation Method},'' \emph{{The
  Annals of Mathematical Statistics}}, vol.~22, no.~3, pp. 400--407, 1951.

\bibitem{tieleman2014rmsprop}
T.~Tieleman and G.~Hinton, ``Rmsprop gradient optimization,'' \emph{URL
  http://www. cs. toronto. edu/tijmen/csc321/slides/lecture\_slides\_lec6.
  pdf}, 2014.

\bibitem{RNN_Elman90findingstructure}
J.~L. Elman, ``Finding structure in time,'' \emph{COGNITIVE SCIENCE}, vol.~14,
  no.~2, pp. 179--211, 1990.

\bibitem{RNN_bengio_grad_in_RNN}
\BIBentryALTinterwordspacing
R.~Pascanu, T.~Mikolov, and Y.~Bengio, ``Understanding the exploding gradient
  problem,'' \emph{CoRR}, vol. abs/1211.5063, 2012. [Online]. Available:
  \url{http://arxiv.org/abs/1211.5063}
\BIBentrySTDinterwordspacing

\bibitem{RNN_TBPTT1}
R.~J. Williams and J.~Peng, ``{An Efficient Gradient-Based Algorithm for
  On-Line Training of Recurrent Network Trajectories},'' \emph{Neural
  Computation}, vol.~2, no.~4, pp. 490--501, Dec 1990.

\bibitem{RNN_TBPTT_sutskever2013}
\BIBentryALTinterwordspacing
I.~Sutskever, ``Training recurrent neural networks,'' \emph{University of
  Toronto, Toronto, Ont., Canada}, 2013. [Online]. Available:
  \url{http://www.cs.utoronto.ca/~ilya/pubs/ilya_sutskever_phd_thesis.pdf}
\BIBentrySTDinterwordspacing

\bibitem{RNN_COlah_LSTM}
\BIBentryALTinterwordspacing
C.~Olah. {Understanding LSTM Networks}. [Online]. Available:
  \url{http://colah.github.io/posts/2015-08-Understanding-LSTMs/}
\BIBentrySTDinterwordspacing

\bibitem{RNN_GRU}
\BIBentryALTinterwordspacing
K.~Cho, B.~van Merrienboer, {\c{C}}.~G{\"{u}}l{\c{c}}ehre, F.~Bougares,
  H.~Schwenk, and Y.~Bengio, ``Learning phrase representations using {RNN}
  encoder-decoder for statistical machine translation,'' \emph{CoRR}, vol.
  abs/1406.1078, 2014. [Online]. Available:
  \url{http://arxiv.org/abs/1406.1078}
\BIBentrySTDinterwordspacing

\bibitem{RNN_schedule-sampl}
\BIBentryALTinterwordspacing
S.~Bengio, O.~Vinyals, N.~Jaitly, and N.~Shazeer, ``{Scheduled Sampling for
  Sequence Prediction with Recurrent Neural Networks},'' \emph{CoRR}, vol.
  abs/1506.03099, 2015. [Online]. Available:
  \url{http://arxiv.org/abs/1506.03099}
\BIBentrySTDinterwordspacing

\bibitem{RNN_coco_img_caption}
\BIBentryALTinterwordspacing
Y.~Cui, M.~R. Ronchi, T.-Y. Lin, P.~Dollr, and L.~Zitnick, ``Microsoft {COCO}
  captioning challenge,'' 2015. [Online]. Available:
  \url{http://cocodataset.org/#captions-challenge2015}
\BIBentrySTDinterwordspacing

\bibitem{RNN_prof_force}
\BIBentryALTinterwordspacing
A.~Goyal, A.~Lamb, Y.~Zhang, S.~Zhang, A.~C. Courville, and Y.~Bengio,
  ``{Professor Forcing: A New Algorithm for Training Recurrent Networks},'' in
  \emph{Advances in Neural Information Processing Systems 29: Annual Conference
  on Neural Information Processing Systems 2016, December 5-10, 2016,
  Barcelona, Spain}, 2016, pp. 4601--4609. [Online]. Available:
  \url{http://papers.nips.cc/paper/6099-professor-forcing-a-new-algorithm-for-training-recurrent-networks}
\BIBentrySTDinterwordspacing

\bibitem{RNN_attention}
\BIBentryALTinterwordspacing
A.~Vaswani, N.~Shazeer, N.~Parmar, J.~Uszkoreit, L.~Jones, A.~N. Gomez,
  L.~Kaiser, and I.~Polosukhin, ``Attention is all you need,'' \emph{CoRR},
  vol. abs/1706.03762, 2017. [Online]. Available:
  \url{http://arxiv.org/abs/1706.03762}
\BIBentrySTDinterwordspacing

\end{thebibliography}

\end{document}